\documentclass[12pt]{article}
\usepackage{amsmath,amssymb}
\usepackage{graphicx}
\newcommand{\eqdef}{\stackrel{\text{def}}{=}}
\newcommand{\eqdefrm}{\stackrel{\text{\rm def}}{=}}
\newcommand{\n}{\nonumber\\}
\newcommand{\bm}{\boldsymbol}
\newcommand{\ignore}[1]{}
\numberwithin{equation}{section}
\newcommand{\Romannumeral}[1]{\uppercase\expandafter{\romannumeral#1}}
\newcommand{\I}{\text{\Romannumeral{1}}}
\newcommand{\II}{\text{\Romannumeral{2}}}
\newtheorem{prop}{\bf Proposition}[section]
\newtheorem{lemma}[prop]{\bf Lemma}

\allowdisplaybreaks[4]

\begin{document}

\baselineskip=20pt

%%%%%%%%%%%%%%%%%%%%%%%%%%%%%%%%%%%%%%%%%%%%%%%%%%%%%%%%%%%%
%                                                          %
%  Title page                                              %
%                                                          %
%%%%%%%%%%%%%%%%%%%%%%%%%%%%%%%%%%%%%%%%%%%%%%%%%%%%%%%%%%%%
\newcommand{\preprint}{
    \begin{flushright}\normalsize \sf
     DPSU-13-3\\
%     {\tt arXiv:1306.nnnn[math-ph]}\\
%     July 2013
   \end{flushright}}
\newcommand{\Title}[1]{{\baselineskip=26pt
   \begin{center} \Large \bf #1 \\ \ \\ \end{center}}}
\newcommand{\Author}{\begin{center}
   \large \bf Satoru Odake${}^a$ and Ryu Sasaki${}^{a,b}$ \end{center}}
\newcommand{\Address}{\begin{center}
     $^a$ Department of Physics, Shinshu University,\\
     Matsumoto 390-8621, Japan\\
     ${}^b$ Center for Theoretical Sciences,\\
     National Taiwan University, Taipei 10617, Taiwan
    \end{center}}
\newcommand{\Accepted}[1]{\begin{center}
   {\large \sf #1}\\ \vspace{1mm}{\small \sf Accepted for Publication}
   \end{center}}

\preprint
\thispagestyle{empty}

\Title{Non-polynomial extensions of solvable potentials\\
\`a la Abraham-Moses}

\Author

\Address
\vspace{1cm}

\begin{abstract}
Abraham-Moses transformations, besides Darboux transformations, are
well-known procedures to generate extensions of solvable potentials in
one-dimensional quantum mechanics. Here we present the explicit forms of
infinitely many {\em seed solutions\/} for adding eigenstates at
{\em arbitrary real energy\/} through the Abraham-Moses transformations
for typical solvable potentials, {\em e.g.\/} the radial oscillator,
the Darboux-P\"oschl-Teller and some others. These seed solutions are simple
generalisations of the {\em virtual state wavefunctions\/}, which are
obtained from the eigenfunctions by discrete symmetries of the potentials.
The virtual state wavefunctions have been an essential ingredient for
constructing {\em multi-indexed\/} Laguerre and Jacobi polynomials
through multiple Darboux-Crum transformations.
In contrast to the Darboux transformations, the virtual state wavefunctions
generate non-polynomial extensions of solvable potentials through the
Abraham-Moses transformations.
\end{abstract}

%%%%%%%%%%%%%%%%%%%%%%%%%%%%%%%%%%%%%%%%%%%%%%%%%%%%%%%%%%%%%%%
%                                                             %
%  1. Introduction                                            %
%                                                             %
%%%%%%%%%%%%%%%%%%%%%%%%%%%%%%%%%%%%%%%%%%%%%%%%%%%%%%%%%%%%%%%
\section{Introduction}
\label{intro}

In order to extend solvable potentials in one-dimensional quantum mechanics
\cite{infhul,susyqm,nieto}, two methods are well-known; the Darboux
transformation \cite{darb,crum} and the Abraham-Moses transformation
\cite{A-M}. The latter, about 30 years old, does not seem to have been
well exploited compared with the former, which is known for about 120 years
and has seen remarkable developments brought about by the multi-indexed
orthogonal polynomials \cite{os25,gomez3} and the exceptional orthogonal
polynomials \cite{os16}--\cite{hos} generated in terms of seed solutions
called {\em virtual state wavefunctions\/}. They are obtained from the
eigenfunctions by the discrete symmetries of the Hamiltonian \cite{os25}.

In this paper, we assert that these virtual state wavefunctions and their
generalisations can also be used for the Abraham-Moses transformations for
adding finitely many eigenstates at {\em arbitrary real energy\/}.
We present the explicit forms of various seed solutions for typical
solvable potentials; the radial oscillator, the Darboux-P\"oschl-Teller
potential and some others. These will bring immensely rich applications of
the Abraham-Moses transformations. The harmonic oscillator potential has
been discussed in the original paper \cite{A-M} and in many others
\cite{AM-rel}--\cite{sam}.

Historically, the Abraham-Moses transformations have been introduced and
discussed in connection with the formulation of the inverse scattering
theory \cite{inv}. However, like the Darboux transformations, as a map
relating one Hamiltonian system to another including the proper solutions,
the Abraham-Moses transformations can be formulated totally algebraically,
without recourse to the inverse scattering theory, so long as the boundary
conditions of various solutions are well specified. The key idea, as
stressed by many authors, is that the Wronskian of two solutions
$\text{W}[\varphi,\psi]$ can be expressed as an integral from one boundary
\eqref{wroneq}; an essential ingredient of the Abraham-Moses transformations.

The present paper is organised as follows.
In section two, the basic formulas of the Abraham-Moses transformations are
recapitulated for introducing necessary notation and for self-containedness.
They are presented algebraically, without making use of the inverse
scattering theory formulation.
Starting from one state adding transformation in \S\,\ref{one}, the multiple
sates addition formulas are given in \S\,\ref{multi}. The multiple states
adding process, starting from a set of $M$ non-normalisable seed solutions
$\{\varphi_j\}$ and ending up as many orthonormal vectors
$(\varphi_j^{(M)},\varphi_k^{(M)})=\delta_{j\,k}$ ($j,k=1,\ldots,M$),
\eqref{varpMform}, can be considered as a good example of an
{\em orthonormalisation procedure of non-normalisable vectors\/}.
The one state deletion is presented in \S\,\ref{del} as an inverse process
of one state addition. The multi-states deletion is commented on briefly.
Various remarks and comments on Abraham-Moses transformations are listed
in \S\,\ref{com} including a note on the relation between Darboux
transformations and Abraham-Moses transformations.
Section three is the main body of the paper. Starting from the two
well-known solvable potentials, the radial oscillator and the
Darboux-P\"oschl-Teller potential, the familiar virtual state wavefunctions
are introduced in \S\,\ref{rsdpt}. They are polynomial type wavefunctions.
For the Darboux-P\"oschl-Teller potential, the total number of addable
eigenstates is limited by the parameters of the starting Hamiltonian.
In \S\,\ref{genvirt} their ``degrees'' are changed to real numbers by
rewriting the Laguerre and Jacobi polynomials as (confluent) hypergeometric
functions. The other genres of seed solutions are also given there.
The seed solutions for other solvable potentials, the Morse potential,
etc, are given in \S\,\ref{other}.
The final section is for a summary and discussions.

%%%%%%%%%%%%%%%%%%%%%%%%%%%%%%%%%%%%%%%%%%%%%%%%%%%%%%%%%%%%%%%
%                                                             %
%  2. Multiple Abraham-Moses Transformations                  %
%                                                             %
%%%%%%%%%%%%%%%%%%%%%%%%%%%%%%%%%%%%%%%%%%%%%%%%%%%%%%%%%%%%%%%
\section{Multiple Abraham-Moses Transformations}
\label{sec:mam}

Here we first recapitulate the essence of the Abraham-Moses transformations
\cite{A-M} for adding one bound state with an {\em arbitrary real\/} energy
in \S\,\ref{one}. By repeating the one state additions, the multiple
Abraham-Moses transformations are realised in \S\,\ref{multi}.
We briefly discuss one and multiple state deletions in \S\,\ref{del}.
The addition and deletion are shown to be the inverse processes of each other.
The other properties are discussed in \S\,\ref{com}.

In contrast to the original and most of the subsequent publications on the
Abraham-Moses transformations \cite{A-M,trl,AM-rel}, our derivations are
purely algebraic without recourse to the inverse scattering method \cite{inv}.
This is partly because some important quantum mechanical systems are defined
in finite intervals, for which the inverse scattering method
is inadequate. The main reason is the clarity of the presentation.
Like the Darboux-Crum transformations \cite{darb,crum,adler}, most
salient features of the Abraham-Moses transformations can be better
understood algebraically.

The starting point is the general quantum mechanics in one dimension defined
in an interval $x_1<x<x_2$ with a smooth potential $U(x)\in\mathbb{R}$.
The system has an infinite (or a finite) number of discrete eigenstates.
For simplicity we assume vanishing groundstate energy:
\begin{align}
  &\mathcal{H}=-\frac{d^2}{dx^2}+U(x),
  \label{schr}\\
  &\mathcal{H}\phi_n(x)=\mathcal{E}_n\phi_n(x)\quad
  (n\in\mathbb{Z}_{\ge0}\ \,\text{or}\ \,0\le n\le n_{\text{max}}),\quad
  0=\mathcal{E}_0<\mathcal{E}_1<\mathcal{E}_2<\cdots,
  \label{sheq}\\
  &(\phi_m,\phi_n)\eqdef\int_{x_1}^{x_2}\!dx\,\phi_m(x)\phi_n(x)
  =h_n\delta_{m\,n},\quad 0<h_n<\infty.
  \label{inpro}
\end{align}
In quantum mechanics, another requirement is built in.
That is, the momentum operator $p=-i\hbar\partial_x$ ($i\equiv\sqrt{-1}$)
must be hermitian. This simply means that the boundary terms
in partial integration should vanish.
We require the following boundary conditions on the eigenfunctions:
\begin{equation}
  \lim_{x\to x_1}\frac{\phi_n(x)^2}{x-x_1}=0,\quad
  \lim_{x\to x_2}\frac{\phi_n(x)^2}{x_2-x}=0\quad(n=0,1,\ldots).
  \label{nbc}
\end{equation}
An appropriate modification is needed when $x_2=+\infty$ and/or $x_1=-\infty$.
Throughout this paper we adopt the convention that all the wavefunctions
are real.
We will not discuss the scattering state wavefunctions.

Let $\{\varphi_j(x),\tilde{\mathcal{E}}_j\}$ ($j=1,2,\ldots,M$) be distinct
solutions of the original Schr\"odinger equation \eqref{schr}:
\begin{equation}
  \mathcal{H}\varphi_j(x)=\tilde{\mathcal{E}}_j\varphi_j(x)\quad
  (\tilde{\mathcal{E}}_j\in\mathbb{R}\ ;\ j=1,2,\ldots,M),
  \label{scheq2}
\end{equation}
to be called {\em seed\/} solutions.
In this paper we consider such seed solutions that are square non-integrable
at one boundary:
\begin{alignat}{3}
  \text{Type $\I$}:\quad&&
  &\lim_{x\to x_1}\frac{\varphi_j(x)^2}{x-x_1}=0,\quad
  &&\int_{x_2-\epsilon}^{x_2}\!\!\!dx\,\varphi_j(x)^2=\infty,
  \label{type1}\\
  \text{Type $\II$}:\quad&&
  &\int_{x_1}^{x_1+\epsilon}\!\!\!dx\,\varphi_j(x)^2=\infty,\quad
  &&\lim_{x\to x_2}\frac{\varphi_j(x)^2}{x_2-x}=0.
  \label{type2}
\end{alignat}
Of course this means that  $\varphi_j(x)$ is square integrable at $x_1$
for type I and at $x_2$ for type II \cite{pursey}.
It should be stressed that there is no type I or type II seed solution
belonging to the spectrum of the Hamiltonian $\{\mathcal{E}_n\}$
($n=0,1,\ldots$), because of the uniqueness of the solutions of the
Schr\"odinger equation.

%%%%%%%%%%%%%%%%%%%%%%%%%%%%%%%%%%%%%%%%%%%%%
%                                           %
% 2.1 One state addition                    %
%                                           %
%%%%%%%%%%%%%%%%%%%%%%%%%%%%%%%%%%%%%%%%%%%%%
\subsection{One state addition}
\label{one}

Let us introduce the Abraham-Moses transformations for {\em adding one
bound state by using a seed solution with an arbitrary real energy\/}.
For simplicity of the presentation, we will restrict ourselves to utilise
the type I seed solutions only. We will comment on the use of the type II
and both I and II in \S\,\ref{com}.
For a pair of real functions $f$ and $g$, which are square integrable
at the lower boundary, let us introduce a new function $\langle f,g\rangle$
by integration:
\begin{align}
  \langle f,g\rangle(x)&\eqdef\int_{x_1}^xdyf(y)g(y)
  =\langle g,f\rangle(x),\quad x_1<x<x_2,\\
  \langle f,g\rangle(x_1)&=0,\quad\langle f,g\rangle(x_2)=(f,g).
\end{align}
Note that $\frac{d}{dx}\langle f,g\rangle(x)=f(x)g(x)$.

For a seed solution, say $\varphi_1$, with the energy $\tilde{\mathcal{E}}_1$,
an Abraham-Moses transformation for adding one bound state with the energy
$\tilde{\mathcal{E}}_1$, is defined as follows:
\begin{align}
  \psi(x)\to\psi^{(1)}(x)
  &\eqdef\psi(x)-\frac{\varphi_1(x)}{1+\langle\varphi_1,\varphi_1\rangle(x)}
  \times\langle\varphi_1,\psi\rangle(x),\\
  \text{or simply}\qquad\qquad\psi\to\psi^{(1)}
  &\eqdef\psi-\frac{\varphi_1}{1+\langle\varphi_1,\varphi_1\rangle}
  \langle\varphi_1,\psi\rangle.
  \label{psi1defs}
\end{align}
Here $\psi$ is an arbitrary smooth function of $x\in(x_1,x_2)$ and
$\langle\varphi_1,\psi\rangle$ must be well defined at the lower boundary
$x_1$.
We have the following:

\begin{prop}\label{prop:one}{\rm \cite{A-M}}
Let $\psi $ be a solution of the original Schr\"odinger equation satisfying
the boundary condition
\begin{equation}
  \mathcal{H}\psi=\mathcal{E}\psi,\ \quad
  \mathcal{E}\in\mathbb{R},\ \quad
  \lim_{x\to x_1}\frac{\psi(x)^2}{x-x_1}=0.
  \label{scheq3}
\end{equation}
Then the function $\psi^{(1)}$ \eqref{psi1defs} satisfies the deformed
Schr\"odinger equation with the same energy:
\begin{align}
  &\mathcal{H}^{(1)}{\psi}^{(1)}=\mathcal{E}{\psi}^{(1)},\\
  &\mathcal{H}^{(1)}\eqdefrm-\frac{d^2}{dx^2}+U^{(1)}(x),\quad
  U^{(1)}(x)\eqdefrm U(x)-2\frac{d^2}{dx^2}
  \log\bigl(1+\langle\varphi_1,\varphi_1\rangle\bigr).
  \label{u1def}
\end{align}
The eigenfunctions are mapped to eigenfunctions with the same norm
\begin{equation}
  \phi_n\to\phi^{(1)}_n,\quad
  (\phi^{(1)}_n,\phi^{(1)}_m)=(\phi_n,\phi_m)=h_n\delta_{n\,m},
  \label{nmort}
\end{equation}
together with the newly created eigenfunction $\varphi_1^{(1)}$, which has
a unit norm:
\begin{equation}
  \varphi_1\to\varphi^{(1)}_1
  =\frac{\varphi_1}{1+\langle\varphi_1,\varphi_1\rangle},\quad
  (\varphi^{(1)}_1,\varphi^{(1)}_1)=1,\quad
  (\varphi^{(1)}_1,\phi^{(1)}_n)=0.
  \label{nvarort}
\end{equation}
\end{prop}

It should be stressed that the seed solution $\varphi_1$ is not square
integrable $(\varphi_1,\varphi_1)=\infty$ and its overall scale is immaterial.
The normalisation part of the Proposition is a simple consequence of the
transformation form \eqref{psi1defs}. The transformed seed solution has
the form:
\begin{equation*}
  \bigl(\varphi^{(1)}_1\bigr)^2
  =\frac{\varphi_1^2}{(1+\langle\varphi_1,\varphi_1\rangle)^2}
  =-\frac{d}{dx}\biggl(\frac{1}{1+\langle\varphi_1,\varphi_1\rangle}\biggr).
\end{equation*}
By integrating the above expression from $x_1$ to $x_2$, we obtain
\begin{equation}
  (\varphi^{(1)}_1,\varphi^{(1)}_1)
  =-\biggl[\frac{1}{1+\langle\varphi_1,\varphi_1\rangle}\biggr]_{x_1}^{x_2}
  =1-\frac{1}{1+(\varphi_1,\varphi_1)}=1.
  \label{varphi11^2}
\end{equation}
Likewise we have
\begin{equation}
  f^{(1)}g^{(1)}=fg-\frac{d}{dx}\biggl(
  \frac{\langle\varphi_1,f\rangle\langle\varphi_1,g\rangle}
  {1+\langle\varphi_1,\varphi_1\rangle}\biggr),
  \label{f1g1}
\end{equation}
for arbitrary smooth functions $f$ and $g$ with well-defined
$\langle\varphi_1,f\rangle$ and $\langle\varphi_1,g\rangle$.
Taking $f=\phi_n$, $g=\phi_m$ and integrating from $x_1$ to $x$, we obtain
\begin{equation}
  \langle\phi^{(1)}_n,\phi^{(1)}_m\rangle
  =\langle\phi_n,\phi_m\rangle
  -\frac{\langle\varphi_1,\phi_n\rangle\langle\varphi_1,\phi_m\rangle}
  {1+\langle\varphi_1,\varphi_1\rangle}.
  \label{intform}
\end{equation}
At the upper boundary $x=x_2$, we obtain \eqref{nmort}.
The other orthogonality relation
%$(\phi^{(1)}_n,\varphi^{(1)}_1)=0$
$(\phi^{(1)}_n,\varphi^{(1)}_1)$ $=0$
\eqref{nvarort} can be shown in a similar way.

Next we note ($f'=\frac{df}{dx}$)
\begin{equation}
  \bigl(\varphi^{(1)}_1\bigr)'
  =\frac{\varphi_1'}{1+\langle\varphi_1,\varphi_1\rangle}
  -\frac{\varphi_1^3}{({1+\langle\varphi_1,\varphi_1\rangle})^2}
  \ \Rightarrow\ \varphi_1\bigl(\varphi^{(1)}_1\bigr)'
  =\varphi_1'\varphi^{(1)}_1-\varphi_1^2\bigl(\varphi^{(1)}_1\bigr)^2,
\end{equation}
which simplifies the expression of the deformed potential
\begin{equation}
  U^{(1)}=U
  -2\biggl(\frac{\varphi_1^2}{1+\langle\varphi_1,\varphi_1\rangle}\biggr)'
  =U-2\bigl(\varphi_1\varphi^{(1)}_1\bigr)'
  =U-2\Bigl(2\varphi_1'\varphi^{(1)}_1
  -\varphi_1^2\bigl(\varphi^{(1)}_1\bigr)^2\Bigr).
\end{equation}
Then it is straightforward to show the deformed Schr\"odinger equation
for ${\varphi}^{(1)}_1$:
\begin{align}
  \bigl(\varphi^{(1)}_1\bigr)''
  &=\frac{\varphi_1''}{1+\langle\varphi_1,\varphi_1\rangle}
  -\frac{4\varphi_1'\varphi_1^2}{(1+\langle\varphi_1,\varphi_1\rangle)^2}
  +\frac{2\varphi_1^5}{(1+\langle\varphi_1,\varphi_1\rangle)^3}\n[4pt]
  &=(U-\tilde{\mathcal{E}}_1){\varphi}^{(1)}_1
  -4\varphi_1'\bigl(\varphi^{(1)}_1\bigr)^2
  +2\varphi_1^2\bigl(\varphi^{(1)}_1\bigr)^3
  =(U^{(1)}-\tilde{\mathcal{E}}_1)\varphi^{(1)}_1.
\end{align}
The deformed Schr\"odinger equation for $\psi^{(1)}$ can be shown as follows:
\begin{align}
  \psi^{(1)}&=\psi-\varphi^{(1)}_1\langle\varphi_1,\psi\rangle
  \ \Rightarrow\ \bigl(\psi^{(1)}\bigr)'
  =\psi'-\bigl(\varphi^{(1)}_1\bigr)'\langle\varphi_1,\psi\rangle
  -\varphi^{(1)}_1\varphi_1\psi,\n
  \bigl(\psi^{(1)}\bigr)''&=\psi''-\bigl(\varphi^{(1)}_1\bigr)''
  \langle\varphi_1,\psi\rangle-2\bigl(\varphi^{(1)}_1\bigr)'\varphi_1\psi
  -\varphi^{(1)}_1(\varphi_1'\psi+\varphi_1\psi')\n
  &=(U-\mathcal{E})\psi
  -(U^{(1)}-\tilde{\mathcal{E}}_1)\varphi^{(1)}_1\langle\varphi_1,\psi\rangle
  -4\varphi_1'\varphi^{(1)}_1\psi
  +2\varphi_1^2\bigl(\varphi^{(1)}_1\bigr)^2\psi\n
  &\ \quad-\varphi^{(1)}_1(\varphi_1\psi'-\varphi_1'\psi)\n
  &=(U^{(1)}-\mathcal{E})\psi^{(1)}-\varphi^{(1)}_1
  \Bigl(\text{W}[\varphi_1,\psi]-(\tilde{\mathcal{E}}_1-\mathcal{E})
  \langle\varphi_1,\psi\rangle\Bigr).
\end{align}
Here $\text{W}[\varphi_1,\psi]$ is the Wronskian,
$\text{W}[\varphi_1,\psi]\eqdef\varphi_1\psi'-\varphi_1'\psi$, satisfying
\begin{equation}
  \bigl(\text{W}[\varphi_1,\psi]\bigr)'
  =(\tilde{\mathcal{E}}_1-\mathcal{E})\varphi_1\psi
 \ \Rightarrow\ \text{W}[\varphi_1,\psi]
  =(\tilde{\mathcal{E}}_1-\mathcal{E})\int_{x_1}^xdy\varphi(y)\psi(y)
  =(\tilde{\mathcal{E}}_1-\mathcal{E})\langle\varphi_1,\psi\rangle,
  \label{wroneq}
\end{equation}
because of the boundary conditions \eqref{nbc}, \eqref{type1} and
\eqref{scheq3}. This proves the deformed Schr\"odinger equation for
$\psi^{(1)}$.

There is one important exceptional situation when
$\mathcal{E}=\tilde{\mathcal{E}}_1$, {\em i.e.\/} at the newly added
eigenenergy level.
In this case, $\text{W}[\varphi_1,\psi]=\text{constant}\neq0$
(otherwise $\psi^{(1)}\propto\varphi^{(1)}_1$) and $\psi^{(1)}$ is no
longer a solution of the deformed Schr\"odinger equation. 

%%%%%%%%%%%%%%%%%%%%%%%%%%%%%%%%%%%%%%%%%%%%%
%                                           %
% 2.2 Multiple states addition              %
%                                           %
%%%%%%%%%%%%%%%%%%%%%%%%%%%%%%%%%%%%%%%%%%%%%
\subsection{Multiple states addition}
\label{multi}

It is now obvious that the new Hamiltonian $\mathcal{H}^{(1)}$ has the
eigenspectrum $\{\tilde{\mathcal{E}}_1,\mathcal{E}_n\}$, and the corresponding
eigenfunctions $\{\varphi^{(1)}_1,\phi_n^{(1)}\}$ ($n=0,1,\ldots$), together
with the seed solution $\{\varphi_j^{(1)}\}$ with energy
$\{\tilde{\mathcal{E}}_j\}$ ($j=2,3,\ldots,M$), assuming that they also
satisfy the boundary conditions.
By picking up another seed solution, say $\varphi_2^{(1)}$, one can define
another Abraham-Moses transformation:
\begin{align*}
  \varphi_2^{(1)}&\to\varphi_2^{(2)}\eqdef
  \frac{\varphi_2^{(1)}}{1+\langle\varphi_2^{(1)},\varphi_2^{(1)}\rangle},\\
  \psi^{(1)}&\to\psi^{(2)}\eqdef
  \psi^{(1)}-\varphi_2^{(2)}\langle\varphi_2^{(1)},\psi^{(1)}\rangle,\\
  \mathcal{H}^{(1)}&\to\mathcal{H}^{(2)}\eqdef
  \mathcal{H}^{(1)}-2\frac{d^2}{dx^2}
  \log\bigl(1+\langle\varphi_2^{(1)},\varphi_2^{(1)}\rangle\bigr),\\
  \mathcal{H}^{(1)}\psi^{(1)}=\mathcal{E}\psi^{(1)}
  &\to\mathcal{H}^{(2)}\psi^{(2)}=\mathcal{E}\psi^{(2)}.
\end{align*}
This step can go on as many as the number of the prepared seed solutions,
so long as the seed functions satisfy the boundary conditions.
Let us use the $M$ seed solutions $\{\varphi_j\}$ \eqref{scheq2} in the order
$j=1,2,\ldots,M$.
At the $K$-th step, the transformation reads:
\begin{align}
  \varphi_K^{(K-1)}&\to\varphi_K^{(K)}\eqdef
  \frac{\varphi_K^{(K-1)}}
  {1+\langle\varphi_K^{(K-1)},\varphi_K^{(K-1)}\rangle},\\
  \psi^{(K-1)}&\to\psi^{(K)}\eqdef
  \psi^{(K-1)}-\varphi_K^{(K)}\langle\varphi_K^{(K-1)},\psi^{(K-1)}\rangle,\\
  \mathcal{H}^{(K-1)}&\to\mathcal{H}^{(K)}\eqdef
  \mathcal{H}^{(K-1)}-2\frac{d^2}{dx^2}
  \log\bigl(1+\langle\varphi_K^{(K-1)},\varphi_K^{(K-1)}\rangle\bigr),\\
  \mathcal{H}^{(K-1)}\psi^{(K-1)}=\mathcal{E}\psi^{(K-1)}
  &\to\mathcal{H}^{(K)}\psi^{(K)}=\mathcal{E}\psi^{(K)},
\end{align}
together with the orthogonality conditions of the eigenfunctions
\begin{align}
  (\phi_n^{(K)},\phi_m^{(K)})&=(\phi_n,\phi_m)=h_n\delta_{n\,m}\quad
  (n,m=0,1,\ldots),
  \label{ortho1}\\
  (\phi_n^{(K)},\varphi_j^{(K)})&=0\quad(n=0,1,\ldots,\,;\,j=1,\ldots,K),\\
  (\varphi_j^{(K)},\varphi_k^{(K)})&=\delta_{j\,k}\quad(j,k=1,\ldots,K).
  \label{varnorm}
\end{align}
The last formula \eqref{varnorm} means that the multiple Abraham-Moses
transformations could be interpreted as {\em orthonormalisation of
non-normalisable vectors\/} $\{\varphi_j\}$.
Indeed the formula \eqref{varnorm} is independent of the fact that the
functions $\{\varphi_j\}$ are the solutions of the Schr\"odinger equation
(cf. \eqref{f1g1}, \eqref{fK-1gK-1}).

The Abraham-Moses transformation for adding one eigenstate
\eqref{psi1defs}--\eqref{u1def} involves one integration. It is naturally
expected that the $K$-fold Abraham-Moses transformation would require
$K$-fold integrals. It turns out that all the higher integrals can be
partially integrated and only simple integrals remain. Let us define an
$M\times M$ symmetric and positive definite matrix $\mathcal{F}$ depending
on the seed solutions $\{\varphi_j\}$ ($j=1,\ldots,M$) as follows:
\begin{equation}
  \mathcal{F}(x)\equiv\mathcal{F}[\varphi_1,\ldots,\varphi_M](x),\quad
  (\mathcal{F})_{j\,k}\eqdef\delta_{j\,k}
  +\langle\varphi_j,\varphi_k\rangle\quad(j,k=1,\ldots,M).
  \label{Fdef}
\end{equation}
For any $M\times M$ matrix $\mathcal{G}$, let us denote by $\mathcal{G}_K$
its $K\times K$ submatrix consisting of $(\mathcal{G})_{j\,k}$
($j,k=1,\ldots,K$). Because of the positive definiteness of $\mathcal{F}_K$,
the inverse $\mathcal{F}_K^{-1}$ is always well-defined.
In terms of $\mathcal{F}_K$ ($K=1,\ldots,M$), we have the following:

\begin{prop}\label{prop:two}{\rm \cite{trl}}
Repeating the one eigenstate adding Abraham-Moses transformations
\eqref{psi1defs}--\eqref{u1def} $M$-times based on the seed solutions
$\{\varphi_j\}$, $j=1,\ldots,M$ in this order, the Hamiltonian
$\mathcal{H}^{(M)}$ and  the corresponding eigenfunctions $\{\phi_n^{(M)}\}$,
$\{\varphi_j^{(M)}\}$ can be expressed in the following simple form:
\begin{align}
  \mathcal{H}^{(M)}&=\mathcal{H}-2\frac{d^2}{dx^2}
  \log\det\bigl(\mathcal{F}_M(x)\bigr),
  \label{Htrans}\\
  \phi^{(M)}_n(x)&=\phi_n(x)-\sum_{j,k=1}^M\varphi_j(x)
  \bigl(\mathcal{F}_M^{-1}(x)\bigr)_{j\,k}\langle\varphi_k,\phi_n\rangle(x)
  \quad(n=0,1,\ldots),
  \label{phiMform}\\
  \varphi^{(M)}_j(x)&=\sum_{k=1}^M
  \bigl(\mathcal{F}^{-1}_M(x)\bigr)_{j\,k}\varphi_k(x),\quad
  (\varphi_j^{(M)},\varphi_k^{(M)})=\delta_{j\,k}\quad(j,k=1,\ldots,M),
  \label{varpMform}
\end{align}
provided that all the intermediate seed solutions satisfy the boundary
conditions.
\end{prop}
Obviously $M=1$ quantities, $\mathcal{H}^{(1)}$ \eqref{u1def},
$\{\phi_n^{(1)}\}$ \eqref{psi1defs} and $\{\varphi_1^{(1)}\}$
\eqref{nvarort} have these forms.

It is rather amusing to verify $M=2$ formulas.
{}From \eqref{intform}, we obtain
\begin{equation*}
  \langle\varphi^{(1)}_2,\varphi^{(1)}_2\rangle
  =\langle\varphi_2,\varphi_2\rangle
  -\frac{\langle\varphi_1,\varphi_2\rangle\langle\varphi_1,\varphi_2\rangle}
  {1+\langle\varphi_1,\varphi_1\rangle},
\end{equation*}
which means that
\begin{align*}
  \bigl(1+\langle\varphi_1,\varphi_1\rangle\bigr)
  \bigl(1+\langle\varphi^{(1)}_2,\varphi^{(1)}_2\rangle\bigr)
  &=\bigl(1+\langle\varphi_1,\varphi_1\rangle\bigr)
  \bigl(1+\langle\varphi_2,\varphi_2\rangle\bigr)
  -\langle\varphi_1,\varphi_2\rangle\langle\varphi_1,\varphi_2\rangle\n
  &=(\mathcal{F}_2)_{1\,1}\times(\mathcal{F}_2)_{2\,2}
  -(\mathcal{F}_2)_{1\,2}\times(\mathcal{F}_2)_{2\,1}=\det(\mathcal{F}_2).
\end{align*}
This proves the potential formula for $U^{(2)}$.
Likewise \eqref{intform} gives
\begin{equation*}
  \langle\varphi^{(1)}_2,\phi^{(1)}_n\rangle
  =\langle\varphi_2,\phi_n\rangle
  -\frac{\langle\varphi_1,\varphi_2\rangle\langle\varphi_1,\phi_n\rangle}
  {1+\langle\varphi_1,\varphi_1\rangle}.
\end{equation*}
This gives an explicit expression of the eigenfunction $\phi_n^{(2)}$ as
a linear combination of terms $\varphi_j\langle\varphi_k,\phi_n\rangle$
($j,k=1,2$) :
\begin{align*}
  \phi_n^{(2)}&=\phi_n^{(1)}
  -\frac{\varphi^{(1)}_2\langle\varphi_2^{(1)},\phi^{(1)}_n\rangle}
  {1+\langle\varphi^{(1)}_2,\varphi^{(1)}_2\rangle}\n
  &=\phi_n-\frac{\varphi_1\langle\varphi_1,\phi_n\rangle}
  {1+\langle\varphi_1,\varphi_1\rangle}\n
  &\quad
  -\biggl(\varphi_2-\frac{\varphi_1\langle\varphi_1,\varphi_2\rangle}
  {1+\langle\varphi_1,\varphi_1\rangle}\biggr)
  \times\biggl(\langle \varphi_2,\phi_n\rangle
  -\frac{\langle\varphi_1,\varphi_2\rangle\langle\varphi_1,\phi_n\rangle}
  {1+\langle\varphi_1,\varphi_1\rangle}\biggr)\times
  \frac1{1+\langle\varphi^{(1)}_2,\varphi^{(1)}_2\rangle}.
\end{align*}
It is indeed trivial to verify that the coefficient of the term
$-\varphi_j\langle\varphi_k,\phi_n\rangle\det(\mathcal{F}_2)^{-1}$ is
the co-factor of the matrix element $(\mathcal{F}_2)_{j\,k}$ ($j,k=1,2$).
This proves the eigenfunction formula \eqref{phiMform} for $M=2$.
The added eigenfunction formula \eqref{varpMform} for $M=2$ can be verified
in a similar manner.

In order to prove Proposition \ref{prop:two} inductively, we need the
following Lemma, with the correspondence $A_n\leftrightarrow \mathcal{F}_K$,
$A_{n-1}\leftrightarrow \mathcal{F}_{K-1}$, $a_{j\,k}\leftrightarrow
\mathcal{F}_{j\,k}=\delta_{j\,k}+\langle\varphi_j,\varphi_k\rangle$.
The Lemma can be proven elementarily by using the cofactor expansion
theorem once or twice.

\begin{lemma}\label{lemm}
For an arbitrary regular matrix $A_n=(a_{j\,k})_{1\le j,k\le n}$ and its
regular submatrix $A_{n-1}=(a_{j\,k})_{1\le j,k\le n-1}$, the following
relations hold
\begin{alignat}{2}
  &({\rm \romannumeral1})&&
  \bigl(A_n^{-1}\bigr)_{n\,n}=\frac{\det(A_{n-1})}{\det(A_n)},
  \label{det1}\\
  &({\rm \romannumeral1}')&&\frac{\det(A_n)}{\det(A_{n-1})}
  =a_{n\,n}-\sum_{j,k=1}^{n-1}a_{n\,j}\bigl(A_{n-1}^{-1}\bigr)_{j\,k}a_{k\,n},
  \label{det2}\\
  &({\rm \romannumeral2})\ 1\le j\le n-1,&&
  \bigl(A_{n-1}^{-1}\bigr)_{j\,n}
  =-\frac{\det(A_{n-1})}{\det(A_n)}
  \sum_{k=1}^{n-1}\bigl(A_{n-1}^{-1}\bigr)_{j\,k}a_{k\,n},
  \label{ainv1}\\
  &({\rm \romannumeral2}')\ 1\le k\le n-1,&&
  \bigl(A_{n-1}^{-1}\bigr)_{n\,k}
  =-\frac{\det(A_{n-1})}{\det(A_n)}
  \sum_{j=1}^{n-1}a_{n\,j}\bigl(A_{n-1}^{-1}\bigr)_{j\,k},
  \label{ainv2}\\
  &({\rm \romannumeral3})\ 1\le j,k\le n-1,\quad&&
  \bigl(A_{n}^{-1}\bigr)_{j\,k}
  =\bigl(A_{n-1}^{-1}\bigr)_{j\,k}+\frac{\det(A_{n-1})}{\det(A_n)}
  \sum_{l,m=1}^{n-1}\bigl(A_{n-1}^{-1}\bigr)_{j\,l}\,a_{l\,n}a_{n\,m}
  \bigl(A_{n-1}^{-1}\bigr)_{m\,k}.
  \label{ainv3}
\end{alignat}
\end{lemma}

Supposing Proposition \ref{prop:two} is true up to $K-1$, we will show
that it is true for $K$.
For an arbitrary smooth function $f$ with well-defined
$\langle\varphi_j,f\rangle$, the $(K-1)$-th transformed function
$f^{(K-1)}$ has the form
\begin{equation}
  f^{(K-1)}=f
  -\sum_{j,k=1}^{K-1}\varphi_j\bigl(\mathcal{F}^{-1}_{K-1}\bigr)_{j\,k}
  \langle\varphi_k,f\rangle.
\end{equation}
For such $f$ and $g$, we have
\begin{align}
  f^{(K-1)}g^{(K-1)}&=fg
  -\sum_{j,k=1}^{K-1}\varphi_jf\bigl(\mathcal{F}^{-1}_{K-1}\bigr)_{j\,k}
  \langle\varphi_k,g\rangle
  -\sum_{j,k=1}^{K-1}\varphi_jg\bigl(\mathcal{F}^{-1}_{K-1}\bigr)_{j\,k}
  \langle\varphi_k,f\rangle\n
  &\quad\ +\!\sum_{j,k,l,m=1}^{K-1}\varphi_j\varphi_l
  \bigl(\mathcal{F}^{-1}_{K-1}\bigr)_{j\,k}
  \bigl(\mathcal{F}^{-1}_{K-1}\bigr)_{l\,m}
  \langle\varphi_k,f\rangle\langle\varphi_m,g\rangle\n
  &=fg-\frac{d}{dx}\biggl(
  \sum_{k,m=1}^{K-1}\langle\varphi_k,f\rangle
  \bigl(\mathcal{F}^{-1}_{K-1}\bigr)_{k\,m}\langle\varphi_m,g\rangle
  \biggr),
  \label{fK-1gK-1}
\end{align}
where we have used
\begin{equation*}
  \sum_{j,l=1}^{K-1}\varphi_j\varphi_l
  \bigl(\mathcal{F}^{-1}_{K-1}\bigr)_{j\,k}
  \bigl(\mathcal{F}^{-1}_{K-1}\bigr)_{l\,m}
  =\sum_{j,l=1}^{K-1}
  \frac{d}{dx}\Bigl((\mathcal{F}_{K-1})_{j\,l}\Bigr)\cdot
  \bigl(\mathcal{F}^{-1}_{K-1}\bigr)_{j\,k}
  \bigl(\mathcal{F}^{-1}_{K-1}\bigr)_{l\,m}
  =-\frac{d}{dx}\bigl(\mathcal{F}^{-1}_{K-1}\bigr)_{k\,m}.
\end{equation*}
Thus we obtain
\begin{equation}
  \langle f^{(K-1)},g^{(K-1)}\rangle
  =\langle f,g\rangle
  -\sum_{j,k=1}^{K-1}\langle\varphi_j,f\rangle
  \bigl(\mathcal{F}^{-1}_{K-1}\bigr)_{j\,k}\langle\varphi_k,g\rangle.
  \label{<fK-1gK-1>}
\end{equation}
The transformation is generated by $\varphi^{(K-1)}_K$,
\begin{equation}
  \varphi^{(K-1)}_K=\varphi_K
  -\sum_{j,k=1}^{K-1}\varphi_j\bigl(\mathcal{F}^{-1}_{K-1}\bigr)_{j\,k}
  \langle\varphi_k,\varphi_K\rangle,
  \label{Mvarphin}
\end{equation}
and \eqref{<fK-1gK-1>} leads to
\begin{align}
  1+\langle\varphi^{(K-1)}_K,\varphi^{(K-1)}_K\rangle
  &=1+\langle\varphi_K,\varphi_K\rangle
  -\sum_{j,k=1}^{K-1}\langle\varphi_j,\varphi_K\rangle
  \bigl(\mathcal{F}^{-1}_{K-1}\bigr)_{j\,k}
  \langle\varphi_k,\varphi_K\rangle\n
%  \label{intK1}\\
%
  &=\frac{\det(\mathcal{F}_K)}{\det(\mathcal{F}_{K-1})}
  =\frac{1}{\bigl(\mathcal{F}_K^{-1}\bigr)_{K\,K}}.
\end{align}
In the second equality, Lemma (\romannumeral1) and (\romannumeral1${}'$)
are used. This proves the change of the potentials \eqref{Htrans} of
Proposition \ref{prop:two}.

Here we introduce a simplifying notation
\begin{equation}
  \alpha\eqdef
  \bigl(1+\langle\varphi^{(K-1)}_K,\varphi^{(K-1)}_K\rangle\bigr)^{-1}
  =\bigl(\mathcal{F}_K^{-1}\bigr)_{K\,K}
  =\frac{\det(\mathcal{F}_{K-1})}{\det(\mathcal{F}_K)}.
\end{equation}
The Abraham-Moses transformation on $\varphi^{(K-1)}_j$ gives for
$1\le j\le K-1$
\begin{equation}
  \varphi^{(K)}_j=\varphi^{(K-1)}_j-\frac{\varphi^{(K-1)}_K
  \langle\varphi^{(K-1)}_K,\varphi^{(K-1)}_j\rangle}
  {1+\langle\varphi^{(K-1)}_K,\varphi^{(K-1)}_K\rangle}.
  \label{K-1K}
\end{equation}
By \eqref{<fK-1gK-1>}, the numerator on the right hand side can be evaluated:
\begin{align}
  \langle\varphi^{(K-1)}_K,\varphi^{(K-1)}_j\rangle
  &=\langle\varphi_K,\varphi_j\rangle
  -\sum_{l,m=1}^{K-1}\langle\varphi_l,\varphi_K\rangle
  \bigl(\mathcal{F}^{-1}_{K-1}\bigr)_{l\,m}\langle\varphi_m,\varphi_j\rangle\n
  &=\sum_{l=1}^{K-1}\langle\varphi_l,\varphi_K\rangle
  \bigl(\mathcal{F}^{-1}_{K-1}\bigr)_{l\,j},
\end{align}
since $\langle\varphi_m,\varphi_j\rangle=(\mathcal{F}_{K-1})_{m\,j}
-\delta_{m\,j}$.
We obtain
\begin{align}
  \varphi^{(K)}_j
  &=\sum_{k=1}^{K-1}\bigl(\mathcal{F}^{-1}_{K-1}\bigr)_{j\,k}\varphi_k
  -\alpha\Bigl(\varphi_K
  -\sum_{k,m=1}^{K-1}\varphi_k\bigl(\mathcal{F}^{-1}_{K-1}\bigr)_{k\,m}
  \langle\varphi_m,\varphi_K\rangle\Bigr)\!
  \times\!\Bigl(\,\sum_{l=1}^{K-1}\langle\varphi_l,\varphi_K\rangle
  \bigl(\mathcal{F}^{-1}_{K-1}\bigr)_{l\,j}\,\Bigr)\n
  &=\sum_{k=1}^{K-1}\varphi_k
  \biggl(\bigl(\mathcal{F}^{-1}_{K-1}\bigr)_{j\,k}
  +\alpha\sum_{l,m=1}^{K-1}\bigl(\mathcal{F}^{-1}_{K-1}\bigr)_{l\,j}
  \bigl(\mathcal{F}^{-1}_{K-1}\bigr)_{k\,m}
  \langle\varphi_l,\varphi_K\rangle\langle\varphi_m,\varphi_K\rangle\biggr)\n
  &\qquad
  -\alpha\,\varphi_K\Bigl(\,\sum_{k=1}^{K-1}\langle\varphi_k,\varphi_K\rangle
  \bigl(\mathcal{F}^{-1}_{K-1}\bigr)_{k\,j}\Bigr).
\end{align}
By Lemma (\romannumeral2) and (\romannumeral3) we arrive at
\begin{equation*} 
  \varphi^{(K)}_j=\sum_{k=1}^K\bigl(\mathcal{F}^{-1}_K\bigr)_{j\,k}\varphi_k
  \quad(j=1,\ldots,K-1).
\end{equation*}
For $j=K$, we obtain directly from \eqref{K-1K},
\begin{equation*}
  \varphi^{(K)}_K=\frac{\varphi^{(K-1)}_K}
  {1+\langle\varphi^{(K-1)}_K,\varphi^{(K-1)}_K\rangle}
  =\alpha\Bigl(\varphi_K-\sum_{j,k=1}^{K-1}\varphi_j
  \bigl(\mathcal{F}^{-1}_{K-1}\bigr)_{j\,k}
  \langle\varphi_k,\varphi_K\rangle\Bigr),
\end{equation*}
which gives the desired result through Lemma (\romannumeral1) and
(\romannumeral2${}'$),
\begin{equation*}
  \varphi^{(K)}_K=\sum_{k=1}^K\bigl(\mathcal{F}^{-1}_K\bigr)_{K\,k}\varphi_k.
\end{equation*}

We apply the Abraham-Moses transformation to $\phi^{(K-1)}_{n}$ by using
the above seed solution:
\begin{equation}
  \phi^{(K)}_n=\phi^{(K-1)}_n
  -\frac{\varphi^{(K-1)}_K\langle\varphi^{(K-1)}_K,\phi^{(K-1)}_n\rangle}
  {1+\langle\varphi^{(K-1)}_K,\varphi^{(K-1)}_K\rangle}.
\end{equation}
{}From \eqref{<fK-1gK-1>}, we have
\begin{equation}
  \langle\varphi^{(K-1)}_K,\phi^{(K-1)}_n\rangle
  =\langle\varphi_K,\phi_n\rangle
  -\sum_{m,k=1}^{K-1}\langle\varphi_m,\varphi_K\rangle
  \bigl(\mathcal{F}^{-1}_{K-1}\bigr)_{m\,k}\langle\varphi_k,\phi_n\rangle.
  \label{Kform}
\end{equation}
We obtain, by using Lemma,
\begin{align*}
  \phi^{(K)}_n&=\phi_n-\sum_{j,k=1}^{K-1}\varphi_j
  \bigl(\mathcal{F}^{-1}_{K-1}\bigr)_{j\,k}\langle\varphi_k,\phi_n\rangle\n
  &\quad-\alpha\Bigl(\varphi_K
  -\sum_{j,l=1}^{K-1}\varphi_j\bigl(\mathcal{F}^{-1}_{K-1}\bigr)_{j\,l}
  \langle\varphi_l,\varphi_K\rangle\Bigr)\!
  \times\!\Bigl(\langle\varphi_K,\phi_n\rangle
  -\sum_{m,k=1}^{K-1}\langle\varphi_m,\varphi_K\rangle
  \bigl(\mathcal{F}^{-1}_{K-1}\bigr)_{m\,k}
  \langle\varphi_k,\phi_n\rangle\Bigr)\n
  &=\phi_n-\biggl(\,
  \sum_{j,k=1}^{K-1}\varphi_j\langle\varphi_k,\phi_n\rangle
  \Bigl(\bigl(\mathcal{F}^{-1}_{K-1}\bigr)_{j\,k}
  +\alpha\sum_{l,m=1}^{K-1}\bigl(\mathcal{F}^{-1}_{K-1}\bigr)_{j\,l}
  \bigl(\mathcal{F}^{-1}_{K-1}\bigr)_{m\,k}
  \langle\varphi_l,\varphi_K\rangle
  \langle\varphi_m,\varphi_K\rangle\Bigr)\n
  &\qquad\qquad\quad
  -\alpha\sum_{j=1}^{K-1}\varphi_j\langle\varphi_K,\phi_n\rangle
  \Bigl(\,\sum_{k=1}^{K-1}\bigl(\mathcal{F}^{-1}_{K-1}\bigr)_{j\,k}
  \langle\varphi_k,\varphi_K\rangle\Bigr)\n
  &\qquad\qquad\quad
  -\alpha\sum_{j=1}^{K-1}\varphi_K\langle\varphi_j,\phi_n\rangle
  \Bigl(\,\sum_{k=1}^{K-1}\bigl(\mathcal{F}^{-1}_{K-1}\bigr)_{k\,j}
  \langle\varphi_k,\varphi_K\rangle\Bigr)
  +\alpha\,\varphi_K\langle\varphi_K,\phi_{n}\rangle\biggr)\n
  &=\phi_n-\sum_{j,k=1}^K\varphi_j\bigl(\mathcal{F}^{-1}_K\bigr)_{j\,k}
  \langle\varphi_k,\phi_n\rangle.
\end{align*}
This concludes the proof of Proposition \ref{prop:two}.
 
It is rather easy to show the orthonormality \eqref{ortho1}--\eqref{varnorm}
based on \eqref{phiMform}--\eqref{varpMform} of Proposition \ref{prop:two}.

%%%%%%%%%%%%%%%%%%%%%%%%%%%%%%%%%%%%%%%%%%%%%
%                                           %
% 2.3 One state deletion                    %
%                                           %
%%%%%%%%%%%%%%%%%%%%%%%%%%%%%%%%%%%%%%%%%%%%%
\subsection{One state deletion}
\label{del}

Deleting multiple eigenstates by Darboux transformation is well established
by Krein-Adler \cite{adler}.
By choosing a subset of the original eigenfunctions \eqref{schr}--\eqref{nbc}
specified by $\mathcal{D}=\{d_1,\ldots,d_M\}$ ($d_j\ge0$), the deleted system
is given by the ratio of Wronskians:
\begin{align*}
  &\text{W}[f_1,f_2,\ldots,f_n](x)\eqdef
  \det\Bigl(\frac{d^{j-1}f_k(x)}{dx^{j-1}}\Bigr)_{1\le j,k\le n},\\
  &\psi\to\psi^{[M]}\eqdef
  \frac{\text{W}[\phi_{d_1},\phi_{d_2},\ldots,\phi_{d_M},\psi]}
  {\text{W}[\phi_{d_1},\phi_{d_2},\ldots,\phi_{d_M}]},\\
  &\phi_n\to\phi_n^{[M]}\eqdef
  \frac{\text{W}[\phi_{d_1},\phi_{d_2},\ldots,\phi_{d_M},\phi_n]}
  {\text{W}[\phi_{d_1},\phi_{d_2},\ldots,\phi_{d_M}]}\quad
  (n=0,1,\ldots,\,;\,n\notin\mathcal{D}),\\[2pt]
  &\mathcal{H}^{[M]}\psi^{[M]}=\mathcal{E}\psi^{[M]},\quad
  \mathcal{H}^{[M]}\phi_n^{[M]}=\mathcal{E}_n\phi_n^{[M]},\\
  &\mathcal{H}^{[M]}\eqdef\mathcal{H}-2\frac{d^2}{dx^2}\log
  \Bigl|\text{W}[\phi_{d_1},\phi_{d_2},\ldots,\phi_{d_M}]\Bigr|,\\
  &(\phi_m^{[M]},\phi_n^{[M]})
  =\prod_{j=1}^M(\mathcal{E}_n-\mathcal{E}_{d_j})\cdot h_n\delta_{m\,n}.
\end{align*}
In order to guarantee the non-singularity of the potential and the positive
definiteness of the norm, the deleted levels must satisfy the conditions
\cite{adler}:
\begin{equation}
  \prod_{j=1}^M(n-d_j)\ge0\quad(\forall n\in\mathbb{Z}_{\ge 0}).
\end{equation}
The conditions mean, in particular, a single state ($M=1$) cannot be deleted
except for the groundstate $\phi_0$. By the Abraham-Moses transformations,
in contrast, one can delete one and many eigenstates.

Here we consider the process of deleting one discrete eigenlevel from the
original Hamiltonian system \eqref{schr}--\eqref{nbc}. Let us denote the
eigenfunction to be deleted by $\phi_d$. By almost the same calculation as
in the case of one state addition, we obtain the following:

\begin{prop}\label{prop:three}{\rm \cite{A-M}}
When the eigenfunction $\phi_d$ has unit norm $(\phi_d,\phi_d)=1$,
the following transformation maps the solution $\psi$ of the original
Hamiltonian system to a solution $\psi^{(1)}$ of the deformed Hamiltonian
system $\mathcal{H}^{(1)}$ \eqref{u1d-def} with the same energy:
\begin{align}
  &\phi_d\to\phi^{(1)}_d\eqdefrm
  \frac{\phi_d}{1-\langle\phi_d,\phi_d\rangle},\quad
  \psi\to\psi^{(1)}\eqdefrm
  \psi+\phi^{(1)}_d\langle\phi_d,\psi\rangle,
  \label{onedel0}\\
  &\phi_n\to\phi^{(1)}_n\eqdefrm
  \phi_n+\phi^{(1)}_d\langle\phi_d,\phi_n\rangle\quad
  (n=0,1,\ldots,\,;\,n\neq d),
  \label{onedel}\\
  &\mathcal{H}^{(1)}\psi^{(1)}=\mathcal{E}\psi^{(1)},\quad
  \mathcal{H}^{(1)}\phi^{(1)}_n=\mathcal{E}_n\phi^{(1)}_n,\\
  &\mathcal{H}^{(1)}\eqdefrm
  -\frac{d^2}{dx^2}+U^{(1)}(x),\quad
  U^{(1)}(x)\eqdefrm U(x)-2\frac{d^2}{dx^2}
  \log\bigl(1-\langle\varphi_1,\varphi_1\rangle\bigr).
  \label{u1d-def}
\end{align}
The norms of the eigenfunctions are preserved except for $\phi^{(1)}_d$,
which becomes {\em non-square integrable\/}.
Thus the eigenstate $\phi_d$ is deleted:
\begin{equation}
  (\phi_n^{(1)},\phi_m^{(1)})=(\phi_n,\phi_m)=h_n\delta_{n\,m}\quad
  (n,m\neq d),\quad 
  (\phi_d^{(1)},\phi_d^{(1)})=\infty.
\end{equation}
\end{prop}

The transformation $\phi_d\to\phi^{(1)}_d$ \eqref{onedel0}--\eqref{u1d-def}
{\em defines a singular Hamiltonian\/}
$\mathcal{H}^{(1)}$, when $\phi_d$  {\em has norm greater than unity\/}
$(\phi_d,\phi_d)>1$. When $\phi_d$'s norm is {\em  less than unity\/}
$(\phi_d,\phi_d)<1$, the new wavefunction $\phi_d^{(1)}$ has a finite
norm and the {\em deletion of the state is not achieved\/}.
If a seed solution $\varphi_d$ is used in the state deleting transformation
\eqref{onedel}, at a certain point $x\in(x_1,x_2)$,
$1-\langle\varphi_d,\varphi_d\rangle$
vanishes and it leads to a singular Hamiltonian.

The fact that the norm of the eigenfunction to be deleted, $\phi_d$, is
strictly restricted to unity can be understood easily when we consider
that the deletion is indeed the inverse process of the addition, in which
all the newly added eigenstates have unit norm, and vice versa.

Let us first add an eigenfunction $\varphi^{(1)}_d$ by using a seed solution
$\varphi_d$, then delete the created eigenfunction $\varphi^{(1)}_d$:
\begin{align*}
  \varphi^{(1)}_d&=\frac{\varphi_d}{1+\langle\varphi_d,\varphi_d\rangle},
  \qquad\qquad
  \varphi^{(2)}_d=\frac{\varphi^{(1)}_d}
  {1-\langle\varphi^{(1)}_d,\varphi^{(1)}_d\rangle},\\
  \phi^{(1)}_n&=\phi_n-\varphi^{(1)}_d\langle\varphi_d,\phi_n\rangle,
  \qquad
  \phi^{(2)}_n=\phi^{(1)}_n
  +\varphi^{(2)}_d\langle\varphi^{(1)}_d,\phi^{(1)}_n\rangle,\\
  \mathcal{H}^{(1)}&=\mathcal{H}-2\frac{d^2}{dx^2}
  \log\bigl(1+\langle\varphi_d,\varphi_d\rangle\bigr),\quad
  \mathcal{H}^{(2)}=\mathcal{H}^{(1)}-2\frac{d^2}{dx^2}
  \log\bigl(1-\langle\varphi^{(1)}_d,\varphi^{(1)}_d\rangle\bigr).
\end{align*}
It is elementary to show (cf. \eqref{varphi11^2}, \eqref{intform})
\begin{align*}
  \langle\varphi^{(1)}_d,\varphi^{(1)}_d\rangle
  &=1-\frac1{1+\langle\varphi_d,\varphi_d\rangle}
  \ \Rightarrow\ \bigl(1+\langle\varphi_d,\varphi_d\rangle\bigr)
  \bigl(1-\langle\varphi^{(1)}_d,\varphi^{(1)}_d\rangle\bigr)=1,\\
  \langle\varphi^{(1)}_d,\phi^{(1)}_n\rangle
  &=\frac{\langle\varphi_d,\phi_n\rangle}{1+\langle\varphi_d,\varphi_d\rangle}.
\end{align*}
These lead, as expected, to:
\begin{align*}
  \varphi^{(2)}_d&=\varphi_d,\quad
  \mathcal{H}^{(2)}=\mathcal{H},\\
  \phi^{(2)}_n&=\phi_n-\varphi^{(1)}_d\langle\varphi_d,\phi_n\rangle
  +\varphi_d\frac{\langle\varphi_d,\phi_n\rangle}
  {1+\langle\varphi_d,\varphi_d\rangle}=\phi_n.
\end{align*}

Next we work in the opposite direction.
We first delete a unit norm eigenstate $\phi_d$, $(\phi_d,\phi_d)=1$,
by mapping it to $\phi^{(1)}_d$, which is not square integrable,
$(\phi^{(1)}_d,\phi^{(1)}_d)=\infty$. Then we add an eigenstate by using
the seed solution $\phi^{(1)}_d$:
\begin{align*}
  \phi^{(1)}_d&=\frac{\phi_d}{1-\langle\phi_d,\phi_d\rangle},
  \qquad\qquad
  \phi^{(2)}_d=\frac{\phi^{(1)}_d}
  {1+\langle\phi^{(1)}_d,\phi^{(1)}_d\rangle},\\
  \phi^{(1)}_n&=\phi_n+\phi^{(1)}_d\langle\phi_d,\phi_n\rangle,
  \qquad
  \phi^{(2)}_n=\phi^{(1)}_n
  -\phi^{(2)}_d\langle\phi^{(1)}_d,\phi^{(1)}_n\rangle,\\
  \mathcal{H}^{(1)}&=\mathcal{H}-2\frac{d^2}{dx^2}
  \log\bigl(1-\langle\phi_d,\phi_d\rangle\bigr),\quad
  \mathcal{H}^{(2)}=\mathcal{H}^{(1)}-2\frac{d^2}{dx^2}
  \log\bigl(1+\langle\phi^{(1)}_d,\phi^{(1)}_d\rangle\bigr).
\end{align*}
It is again elementary to show
\begin{align*}
  \langle\phi^{(1)}_d,\phi^{(1)}_d\rangle
  &=\frac1{1-\langle\phi_d,\phi_d\rangle}-1
  \ \Rightarrow\ \bigl(1-\langle\phi_d,\phi_d\rangle\bigr)
  \bigl(1+\langle\phi^{(1)}_d,\phi^{(1)}_d\rangle\bigr)=1,\\
  \langle\phi^{(1)}_d,\phi^{(1)}_n\rangle
  &=\frac{\langle\phi_d,\phi_n\rangle}{1-\langle\phi_d,\phi_d\rangle}.
\end{align*}
These lead, as expected, to:
\begin{align*}
  \phi^{(2)}_d&=\phi_d,\quad
  \mathcal{H}^{(2)}=\mathcal{H},\\
  \phi^{(2)}_n&=\phi_n+\phi^{(1)}_d\langle\phi_d,\phi_n\rangle
  -\phi_d\frac{\langle\phi_d,\phi_n\rangle}
  {1-\langle\phi_d,\phi_d\rangle}=\phi_n.
\end{align*}

At the end of this subsection let us present the formulas of multiple
eigenstates deletion by using $M$ eigenfunctions $\{\phi_{d_j}\}$,
$(\phi_{d_j},\phi_{d_k})=\delta_{j\,k}$, $j=1,\ldots,M$, in this order:
\begin{align}
  \mathcal{H}^{(M)}&=\mathcal{H}
  -2\frac{d^2}{dx^2}\log\det\bigl(\bar{\mathcal{F}}_M(x)\bigr),\quad
  \mathcal{D}\eqdef\{d_1,\ldots,d_M\},
  \label{dHtrans}\\
  \phi^{(M)}_n(x)&=\phi_n(x)
  +\sum_{j,k=1}^M\phi_{d_j}(x)\bigl(\bar{\mathcal{F}}_M^{-1}(x)\bigr)_{j\,k}
  \langle\phi_{d_k},\phi_n\rangle(x)\quad
  (n=0,1,\ldots,\,;\,n\notin\mathcal{D}),
  \label{dphiMform}\\
  \phi^{(M)}_{d_j}(x)&
  =\sum_{k=1}^M\bigl(\bar{\mathcal{F}}^{-1}_M(x)\bigr)_{j\,k}\phi_{d_k}(x)
  \quad(j=1,\ldots,M),
  \label{dvarpMform}\\
  \bar{\mathcal{F}}(x)&\equiv
  \bar{\mathcal{F}}[\phi_{d_1},\ldots,\phi_{d_M}](x)
  =(\bar{\mathcal{F}}_{j,k})_{1\leq j,k\leq M},\quad
  (\bar{\mathcal{F}})_{j\,k}
  \eqdef\delta_{j\,k}-\langle\phi_{d_j},\phi_{d_k}\rangle.
\end{align}
These formulas are almost the same as those for the multiple eigenstate
addition \eqref{Htrans}--\eqref{varpMform} in Proposition \ref{prop:two},
with $\mathcal{F}$ replaced by $\bar{\mathcal{F}}$ and a plus sign in
\eqref{dphiMform} instead of a minus sign in \eqref{phiMform}.
The proof goes parallel with the multiple eigenstate addition case.
Indeed these formulas are obtained from those for the multiple eigenstate
addition \eqref{Htrans}--\eqref{varpMform} by changing
$\varphi_j\to i\phi_{d_j}$, $\psi\to i\psi$ and
$\phi_n\to i\phi_n$ ($n\notin\mathcal{D}$), $i\equiv\sqrt{-1}$.

%%%%%%%%%%%%%%%%%%%%%%%%%%%%%%%%%%%%%%%%%%%%%%%%%
%                                               %
% 2.4 Comments on Abraham-Moses transformations %
%                                               %
%%%%%%%%%%%%%%%%%%%%%%%%%%%%%%%%%%%%%%%%%%%%%%%%%
\subsection{Comments on Abraham-Moses transformations}
\label{com}

Here are some comments on various aspects of the Abraham-Moses transformations.
As for the seed solutions for adding eigenstates \eqref{scheq2}, we have
not specified the overall scale of these functions, since there is no
standard way of fixing the scale of such non square integrable functions.
The very fact that the obtained eigenfunctions have unit norms is independent
of such overall scales.
As stressed in Abraham-Moses paper \cite{A-M}, one could use one of the
original eigenfunctions, $\phi_a$ with energy $\mathcal{E}_a$, as a seed
solution. In this case, $\phi_a^{(1)}=\phi_a/(1+\langle\phi_a,\phi_a\rangle)$
is still an eigenfunction with energy $\mathcal{E}_a$. Its norm is changed
to $(\phi_a^{(1)},\phi_a^{(1)})=(\phi_a,\phi_a)/(1+(\phi_a,\phi_a))$.

As for the type II seed solutions \eqref{type2} \cite{pursey}, we have to
change the definition of the function $\langle f,g\rangle(x)$ as follows:
\begin{align}
  \langle f,g\rangle(x)&\eqdef-\int_{x_2}^xdyf(y)g(y)
  =\langle g,f\rangle(x),\quad x_1<x<x_2,\\
  \langle f,g\rangle(x_1)&=(f,g),\quad \langle f,g\rangle(x_2)=0.
\end{align}
Then all the formulas in this section are also true when the type II
seed solutions only are used.

It is definitely true that one can apply the state adding Abraham-Moses
transformations in terms of both type I and II seed solutions in any order,
if seed solutions of one type remain seed solutions after transformations
by the other type. This depends on the explicit forms of the seed solutions.
Let us consider a seed solution $\varphi_2$ of type II after the
transformation by a seed solution $\varphi_1$ of type I:
\begin{equation*}
  \varphi^{(1)}_2=\varphi_2-\varphi^{(1)}_1\langle\varphi_1,\varphi_2\rangle.
\end{equation*}
By construction
$\varphi^{(1)}_1 =\varphi_1/(1+\langle\varphi_1,\varphi_1\rangle)$ is
well behaved on both boundaries.
If the integral $\langle\varphi_1,\varphi_2\rangle(x)
=\int_{x_1}^xdy\varphi_1(y)\varphi_2(y)$ exists on both boundaries or
its certain regularisation exists, it is highly likely that
$\varphi^{(1)}_2$ can qualify as a type II seed solution.
The situation is about the same for a seed solution of type I after the
transformation by a type II seed solution.

Even when these mixed multiple transformations are possible, to write down
the generic formulas like Proposition \ref{prop:two} for such Abraham-Moses
transformations is a different matter.
In contrast to the multiple Darboux transformations in terms of type I and
II virtual state wavefunctions worked out for the radial oscillator,
Darboux-P\"oschl-Teller and other solvable potentials \cite{os25,os28,os29},
we are not quite sure if generic formulas exist for the multiple state
adding Abraham-Moses transformations in terms of both type I and II seed
solutions.

Let us briefly comment on the relation between a Darboux transformation
and one state adding Abraham-Moses transformation \cite{schn-leeb,sam}.
Let us first execute a Darboux transformation by picking up a seed solution
$(\varphi,\tilde{\mathcal{E}})$ of type I \eqref{type1}:
\begin{align*}
  \mathcal{H}\to\mathcal{H}^{[1]}&\eqdef
  -\frac{d^2}{dx^2}+U^{[1]}(x),\quad
  U^{[1]}(x)\eqdef U(x)-2\frac{d^2}{dx^2}\log|\varphi|,\\
  \psi\to\psi^{[1]}&\eqdef
  \psi'-\frac{\varphi'}{\varphi}\psi,\quad
  \mathcal{H}^{[1]}\psi^{[1]}=\mathcal{E}\psi^{[1]}.
\end{align*}
Next we perform a second Darboux transformation in terms of a particular
solution of $\mathcal{H}^{[1]}$:
\begin{align}
  \bar{\varphi}^{[1]}
  &\eqdef\frac1{\varphi}(1+\langle\varphi,\varphi\rangle),\quad 
  \mathcal{H}^{[1]}\bar{\varphi}^{[1]}=\tilde{\mathcal{E}}\bar{\varphi}^{[1]},
  \label{barphi}\\
  \mathcal{H}^{[1]}\to\mathcal{H}^{[2]}
  &\eqdef\mathcal{H}^{[1]}-2\frac{d^2}{dx^2}
  \log\bigl|\,\bar{\varphi}^{[1]}\bigr|
  =\mathcal{H}-2\frac{d^2}{dx^2}
  \log\bigl(1+\langle\varphi,\varphi\rangle\bigr),\n
  \psi^{[1]}\to\psi^{[2]}
  &\eqdef\bigl(\psi^{[1]}\bigr)'
  -\frac{(\bar{\varphi}^{[1]})'}{\bar{\varphi}^{[1]}}\psi^{[1]}
  =(\tilde{\mathcal{E}}-\mathcal{E})\psi-\frac{1}{\bar{\varphi}^{[1]}}
  \text{W}[\varphi,\psi].\nonumber
\end{align}
Since $\psi$ and $\varphi$ satisfy the boundary conditions \eqref{nbc} and
\eqref{type1}, the Wronskian $\text{W}[\varphi,\psi]$ can be expressed in
terms of an integral $\text{W}[\varphi,\psi]=(\tilde{\mathcal{E}}
-\mathcal{E})\langle\varphi,\psi\rangle$ as in \eqref{wroneq}.
Thus we arrive at the one state adding Abraham-Moses transformation
\eqref{psi1defs}
\begin{equation}
  \psi^{[2]}=(\tilde{\mathcal{E}}-\mathcal{E})
  \Bigl(\psi-\frac{\varphi}{1+\langle\varphi,\varphi\rangle}
  \langle\varphi,\psi\rangle\Bigr).
\end{equation}
Instead of a seed solution ($\varphi$, $\tilde{\mathcal{E}}$), an eigenstate
($\phi_d$, $\mathcal{E}_d$) and
$\bar{\phi}_d^{[1]}\eqdef\frac{1}{\phi_d}(1-\langle\phi_d,\phi_d\rangle)$
are used, the one eigenstate deleting Abraham-Moses transformation
\eqref{onedel} is obtained. The relation between the two seed solutions
$\varphi\leftrightarrow\bar{\varphi}^{[1]}$ \eqref{barphi} and its many
disguises have been discussed by many authors in connection with Abraham-Moses
transformations \cite{AM-rel}.

%%%%%%%%%%%%%%%%%%%%%%%%%%%%%%%%%%%%%%%%%%%%%%%%%%%%%%%%%%%%%%%
%                                                             %
%  3. Generalised Virtual State Wavefunctions                 %
%                                                             %
%%%%%%%%%%%%%%%%%%%%%%%%%%%%%%%%%%%%%%%%%%%%%%%%%%%%%%%%%%%%%%%
\section{Generalised Virtual State Wavefunctions}
\label{sec:gen}

Here we present the explicit forms of various seed solutions for some
exactly solvable potentials \cite{infhul,susyqm}, in particular the
radial oscillator and the Darboux-P\"oschl-Teller potential and a few more.
(The harmonic oscillator case has been discussed in the original
Abraham-Moses paper \cite{A-M}.)
They are necessary in order to carry out the program of `generating exactly
solvable potentials' by adding {\em a finite number of eigenstates\/} with
{\em arbitrary energies\/} through multiple Abraham-Moses transformations.
These seed solutions are tentatively called `generalised virtual state
wavefunctions.' The `virtual state wavefunctions' have been introduced by
the present authors \cite{os25,os16,os19} and extensively used to generate
`rational or polynomial extensions' of various solvable potentials,
through multiple Darboux-Crum transformations \cite{darb,crum,adler}.
Obtained from the eigenfunctions by a discrete symmetry operation of the
original Hamiltonian, these virtual state wavefunctions are of polynomial
character and their energies are discretised and negative by restricting
the ranges of their degrees. They have been indispensable for the
construction of the `multi-indexed Jacobi and Laguerre polynomials'
\cite{os25} including various exceptional orthogonal polynomials as the
simplest cases \cite{gomez}--\cite{hos}. Since the negative energy condition
is irrelevant, these virtual state wavefunctions of type I and II, without
any restrictions to their degrees, are bona fide seed solutions for
Abraham-Moses transformations, easiest to use in practical applications.

In order to construct {\em seed solutions of arbitrary real energies\/},
we generalise the polynomial type virtual state wavefunctions as well as
the eigenfunctions to hypergeometric functions (${}_2F_1$ and ${}_1F_1$) type,
by making the degree of polynomial type solutions to be a continuous real
number. This has been done in our previous paper \cite{os21}.
See also \cite{junkroy}.
For the Darboux-P\"oschl-Teller potential, we also report another genre of
real seed solutions corresponding to `complex degrees'.

%%%%%%%%%%%%%%%%%%%%%%%%%%%%%%%%%%%%%%%%%%%%%%%%%%%%%%%%%%%%%%%%
%                                                              %
% 3.1 Radial oscillator and Darboux-P\"oschl-Tellet potentials %
%                                                              %
%%%%%%%%%%%%%%%%%%%%%%%%%%%%%%%%%%%%%%%%%%%%%%%%%%%%%%%%%%%%%%%%
\subsection{Radial oscillator and Darboux-P\"oschl-Teller potentials}
\label{rsdpt}

We first recapitulate the known virtual state wavefunctions of type I and
II of the Hamiltonian systems with the radial oscillator and
Darboux-P\"oschl-Teller potentials. The potentials are
\begin{equation}
  U(x)=\left\{
  \begin{array}{llll}
  {\displaystyle x^2+\frac{g(g-1)}{x^2}-(1+2g)},&x_1=0,\ x_2=\infty,
  &g>\frac32&:\text{L}\\[5pt]
  {\displaystyle \frac{g(g-1)}{\sin^2x}+\frac{h(h-1)}{\cos^2x}-(g+h)^2},
  &x_1=0,\ x_2=\frac{\pi}{2},& g,\,h>\frac32&:\text{J}
  \end{array}\right.,
  \label{pots}
\end{equation}
in which L and J stand for the names of their eigenfunctions, the Laguerre
and Jacobi polynomials. The parameter ranges are consistent with the
restrictions for the eigenfunctions \eqref{nbc} and for the seed solutions
\eqref{type1}, \eqref{type2}.
The eigenfunctions are factorised into the groundstate eigenfunction and
the polynomial in a functions $\eta=\eta(x)$, called {\em sinusoidal
coordinate\/} \cite{nieto,os7}:
\begin{equation}
  \phi_n(x;\bm{\lambda})=\phi_0(x;\bm{\lambda})
  P_n\bigl(\eta(x);\bm{\lambda}\bigr),
  \label{factor}
\end{equation}
in which $\bm{\lambda}$ stands for the parameters, $g$ for L and $(g,h)$
for J. Their explicit forms are
\begin{align}
  \text{L}:\quad&\phi_0(x;g)=e^{-\frac12x^2}x^g,\quad
  P_n(\eta;g)=L_n^{(g-\frac12)}(\eta),\quad\eta(x)=x^2,\n
  &\mathcal{E}_n(g)=4n,\quad
  h_n(g)=\frac{1}{2\,n!}\Gamma(n+g+\tfrac12),
  \label{phi0L} \\
  \text{J}:\quad&\phi_0(x;g,h)=(\sin x)^g(\cos x)^h,\quad
  P_n(\eta;g,h)=P_n^{(g-\frac12,h-\frac12)}(\eta),\quad
  \eta(x)=\cos2x,\n
  &\mathcal{E}_n(g,h)=4n(n+g+h),\quad
  h_n(g,h)=\frac{\Gamma(n+g+\frac12)\Gamma(n+h+\frac12)}
  {2\,n!(2n+g+h)\Gamma(n+g+h)},
  \label{phi0J}
\end{align}
in which $h_n$ is the normalisation constant of the norm introduced in
\eqref{inpro}.

It is obvious that the above potential \eqref{pots} without the constant
term ($-(1+2g)$ for L and $-(g+h)^2$ for J) are invariant under the
discrete transformation $g\leftrightarrow 1-g$ and/or $h\leftrightarrow 1-h$.
The Hamiltonian for L without the constant term changes the sign under
$x\to ix$. These are the discrete symmetry transformations mapping the
above eigenfunctions to seed solutions of type I and II, which are again
{\em polynomial solutions\/}.
The virtual states wavefunctions for L are:
\begin{align}
  \text{L1}:\quad
  &\tilde{\phi}_\text{v}^{\I}(x;g)\eqdef
  e^{\frac12x^2}x^gL_\text{v}^{(g-\frac12)}\bigl(-\eta(x)\bigr),\quad\quad
  \tilde{\mathcal{E}}_\text{v}^{\I}(g)=-4(g+\text{v}+\tfrac12)\quad
  (\text{v}\in\mathbb{Z}_{\ge0}),
  \label{vsL1}\\
  \text{L2}:\quad
  &\tilde{\phi}_\text{v}^{\II}(x;g)\eqdef
  e^{-\frac12x^2}x^{1-g}L_\text{v}^{(\frac12-g)}\bigl(\eta(x)\bigr),\quad
  \tilde{\mathcal{E}}_\text{v}^{\II}(g)=-4(g-\text{v}-\tfrac12)\quad
  (\text{v}\in\mathbb{Z}_{\ge0}).
  \label{vsL2}
\end{align}
The virtual states wavefunctions for J are:
\begin{align}
  \text{J1}:\quad
  &\tilde{\phi}_\text{v}^{\I}(x;g,h)\eqdef(\sin x)^g(\cos x)^{1-h}
  P_\text{v}^{(g-\frac12,\frac12-h)}\bigl(\eta(x)\bigr),\n
  &\tilde{\mathcal{E}}_\text{v}^{\I}(g,h)=-4(g+\text{v}+\tfrac12)
  (h-\text{v}-\tfrac12)\quad
  (\text{v}\in\mathbb{Z}_{\ge0}),
  \label{vsJ1}\\
  \text{J2}:\quad
  &\tilde{\phi}_\text{v}^{\II}(x;g,h)\eqdef(\sin x)^{1-g}(\cos x)^h
  P_\text{v}^{(\frac12-g,h-\frac12)}\bigl(\eta(x)\bigr),\n
  &\tilde{\mathcal{E}}_\text{v}^{\II}(g,h)=-4(g-\text{v}-\tfrac12)
  (h+\text{v}+\tfrac12)\quad
  (\text{v}\in\mathbb{Z}_{\ge0}).
  \label{vsJ2}
\end{align}
Due to the parity property of the Jacobi polynomial
$P_n^{(\alpha,\beta)}(-x)=(-1)^nP_n^{(\beta,\alpha)}(x)$, the two virtual
state polynomials for J are related by this parity transformation. It is
obvious that the virtual state wavefunctions satisfy the boundary
conditions \eqref{type1} and \eqref{type2} and that the type I solutions
are not square integrable at the upper boundary $x_2$ and the type II
solutions are not square integrable at the lower boundary $x_1$.
If the two types of the discrete symmetry operations are applied, the
resulting solutions are not square integrable at either boundary.
They are called pseudo virtual state wavefunctions \cite{os29} and they
cannot be used for the Abraham-Moses transformations.

At each step of state adding Abraham-Moses transformation, the parameters
of the above virtual state wavefunctions describing the boundary conditions
change:
\begin{equation}
  \text{J1}:\ \ h\to h-2,\qquad\text{L2 \& J2}:\ \ g\to g-2.
\end{equation}
These are consistent with the interpretation that Abraham-Moses
transformations can be understood as special two-step Darboux transformations.
This also means that the total number of addable eigenstates is limited
when using the above J1, J2 and L2 seed solutions. As stressed in \cite{os19},
the L1 case is obtained from J1 by the confluence limit, $h\to\infty$.
Thus their boundary conditions are not affected by each Abraham-Moses
transformation and the L1 virtual state wavefunctions can be used as many
as wanted.

These known virtual state wavefunctions are all of polynomial type and
their energies $\tilde{\mathcal{E}}_{\text v}$ take only discretised
values for integer $\text{v}$, which is the degree of the polynomial.
In the next subsection, we generalise the virtual state wavefunctions
to take arbitrary real energies.
It should be easy for each explicit example of seed solutions to calculate
the change of the boundary parameters as above.

%%%%%%%%%%%%%%%%%%%%%%%%%%%%%%%%%%%%%%%%%%%%%%%%%%
%                                                %
% 3.1.1. generalised virtual state wavefunctions %
%                                                %
%%%%%%%%%%%%%%%%%%%%%%%%%%%%%%%%%%%%%%%%%%%%%%%%%%
\subsubsection{generalised virtual state wavefunctions}
\label{genvirt}

The strategy for the generalisation is quite simple, as shown in
\cite{os21} for the type I cases. We rewrite the Laguerre and Jacobi
polynomials in terms of (confluent) hypergeometric functions:
\begin{align}
  L^{(\alpha)}_n(x)&=\frac{(\alpha+1)_n}{n!}
  \sum_{k=0}^n\frac{(-n)_k}{(\alpha+1)_k}\frac{x^k}{k!}\n
  &=\frac{\Gamma(\alpha+1+n)}{\Gamma(\alpha+1)\Gamma(n+1)}\,
  {}_1F_1\Bigl(\genfrac{}{}{0pt}{}{-n}{\alpha+1}\Bigm|x\Bigr),\\
  P^{(\alpha,\beta)}_n(x)&=\frac{(\alpha+1)_n}{n!}
  \sum_{k=0}^n\frac{(-n)_k(n+\alpha+\beta+1)_k}
  {(\alpha+1)_k\,k!}\Bigl(\frac{1-x}{2}\Bigr)^k\n
  &=\frac{\Gamma(\alpha+1+n)}{\Gamma(\alpha+1)\Gamma(n+1)}\,
  {}_2F_1\Bigl(\genfrac{}{}{0pt}{}{-n,n+\alpha+\beta+1}
  {\alpha+1}\Bigm|\frac{1-x}{2}\Bigr).
\end{align}
The expressions in terms of (confluent) hypergeometric functions are valid
for any complex number $n$ and satisfy the Laguerre and Jacobi's differential
equation, respectively.
Since the overall scale of the seed solutions is irrelevant, we will drop
the overall factors.
The non-polynomial forms of the seed solutions valid for a real number
$\text{v}$ ($\text{v}\in\mathbb{R}$, $\text{v}\not\in\mathbb{Z}_{\geq 0}$) are
\begin{align}
  \text{L1}:\quad
  &\tilde{\phi}_\text{v}^{\I}(x;g)\eqdef
  e^{\frac12x^2}x^g
  {}_1F_1\Bigl(\genfrac{}{}{0pt}{}{-\text{v}}{g+\frac12}\Bigm|-\eta(x)\Bigr),
  \label{L1gen}\\
  \text{L2}:\quad
  &\tilde{\phi}_\text{v}^{\II}(x;g)\eqdef
  e^{-\frac12x^2}x^{1-g}
  {}_1F_1\Bigl(\genfrac{}{}{0pt}{}{-\text{v}}{\frac32-g}\Bigm|\eta(x)\Bigr),
  \label{L2gen}\\
  \text{J1}:\quad
  &\tilde{\phi}_\text{v}^{\I}(x;g,h)\eqdef(\sin x)^g(\cos x)^{1-h}
  {}_2F_1\Bigl(\genfrac{}{}{0pt}{}{-\text{v},\text{v}+g-h+1}
  {g+\frac12}\Bigm|\frac{1-\eta(x)}{2}\Bigr),
  \label{J1gen}\\
  \text{J2}:\quad
  &\tilde{\phi}_\text{v}^{\II}(x;g,h)\eqdef(\sin x)^{1-g}(\cos x)^h
  {}_2F_1\Bigl(\genfrac{}{}{0pt}{}{-\text{v},\text{v}+h-g+1}
  {h+\frac12}\Bigm|\frac{1+\eta(x)}{2}\Bigr).
  \label{J2gen}
\end{align}
The energy formulas \eqref{vsL1}--\eqref{vsJ2} are now valid for any real
number $\text{v}$.
By using the Kummer's transformation formulas,
\begin{align}
  {}_1F_1\Bigl(\genfrac{}{}{0pt}{}{\alpha}
  {\beta}\Bigm|x\Bigr)&=e^{x}{}_1F_1\Bigl(\genfrac{}{}{0pt}{}{\beta-\alpha}
  {\beta}\Bigm|-x\Bigr),\\
  {}_2F_1\Bigl(\genfrac{}{}{0pt}{}{\alpha,\beta}
  {\gamma}\Bigm|x\Bigr)&=(1-x)^{\gamma-\alpha-\beta}
  {}_2F_1\Bigl(\genfrac{}{}{0pt}{}{\gamma-\alpha,\gamma-\beta}
  {\gamma}\Bigm|x\Bigr),
\end{align}
they can be rewritten \cite{os21}. For example,
\begin{align*}
  \text{L1}\ \eqref{L1gen}\quad&
  \tilde{\phi}_\text{v}^{\I}(x;g)=
  e^{-\frac12x^2}x^g
  {}_1F_1\Bigl(\genfrac{}{}{0pt}{}{g+\frac12+\text{v}}{g+\frac12}
  \Bigm|\eta(x)\Bigr),\\
  \text{J1}\ \eqref{J1gen}\quad&
  \tilde{\phi}_\text{v}^{\I}(x;g,h)=(\sin x)^g(\cos x)^h
  {}_2F_1\Bigl(\genfrac{}{}{0pt}{}{g+\frac12+\text{v},h-\frac12-\text{v}}
  {g+\frac12}\Bigm|\frac{1-\eta(x)}{2}\Bigr).
\end{align*}

The same procedure, {\em polynomials to (confluent) hypergeometric series\/},
can be applied to the eigenfunctions to obtain another type of L1, J1 and
J2 seed solutions ($\text{v}\in\mathbb{R}$,
$\text{v}\not\in\mathbb{Z}_{\geq 0}$) :
\begin{align}
  \text{L1}:\quad
  &\phi_{\text{v}}^{\I}(x;g)\eqdef e^{-\frac12x^2}x^g
  {}_1F_1\Bigl(\genfrac{}{}{0pt}{}{-\text{v}}{g+\frac12}\Bigm|\eta(x)\Bigr),
  \label{L1eigen}\\
  \text{J1}:\quad
  &\phi_{\text{v}}^{\I}(x;g,h)\eqdef(\sin x)^g(\cos x)^h
  {}_2F_1\Bigl(\genfrac{}{}{0pt}{}{-\text{v},\text{v}+g+h}
  {g+\frac12}\Bigm|\frac{1-\eta(x)}{2}\Bigr),
  \label{J1eigen}\\
  \text{J2}:\quad
  &\phi_{\text{v}}^{\II}(x;g,h)\eqdef(\sin x)^g(\cos x)^h
  {}_2F_1\Bigl(\genfrac{}{}{0pt}{}{-\text{v},\text{v}+g+h}
  {h+\frac12}\Bigm|\frac{1+\eta(x)}{2}\Bigr).
  \label{J2eigen}
\end{align}

Another generalisation exists for the J1, J2 seed solutions.
For certain complex values of $\text{v}$ ($B\in\mathbb{R}$, $B\neq 0$) :
\begin{align}
  &\left\{
  \begin{array}{ll}
  \text{J1}\ \eqref{J1gen}:&\text{v}=\frac12(h-g-1)+iB\\[2pt]
  \text{J2}\ \eqref{J2gen}:&\text{v}=\frac12(g-h-1)+iB
  \end{array}\right.,\\
  &\left\{
  \begin{array}{ll}
  \text{J1}\ \eqref{J1eigen}:&\text{v}=-\frac12(g+h)+iB\\[2pt]
  \text{J2}\ \eqref{J2eigen}:&\text{v}=-\frac12(g+h)+iB
  \end{array}\right.,
\end{align}
the seed solutions \eqref{J1gen}, \eqref{J2gen}, \eqref{J1eigen} and
\eqref{J2eigen} and the corresponding energies \eqref{phi0J}, \eqref{vsJ1}
and \eqref{vsJ2} are real:
\begin{equation}
  \tilde{\mathcal{E}}_\text{v}(g,h)=-\bigl((g+h)^2+4B^2\bigr)<0.
\end{equation}

%%%%%%%%%%%%%%%%%%%%%%%%%%%%%%%%%%%%%%%%%%%%%%
%                                            %
% 3.2. Other solvable potentials             %
%                                            %
%%%%%%%%%%%%%%%%%%%%%%%%%%%%%%%%%%%%%%%%%%%%%%
\subsection{Other solvable potentials }
\label{other}

Among various solvable potentials, some have only {\em finitely many
discrete eigenstates\/}, which are labeled by the degrees of the polynomial
eigenfunctions, $n=0,1,\ldots,n_{\text{max}}$.
For these, the same polynomial wavefunctions as the eigenfunctions with
higher degrees than the highest energy eigenfunction $n>n_{\text{max}}$
provide seed solutions, on the assumption that the boundary condition at
one boundary is satisfied \cite{quesne4,quesne5,grandati2}. These are called
{\em overshoot eigenfunctions\/}, \cite{os28}. Here we report that the
following four potentials have overshoot eigenfunctions.
 
%%%%%%%%%%%%%%%%%%%%%%%%%%%%%%%%%%%%%%%%%%%%%%
%                                            %
% 3.2.1. Morse potential                     %
%                                            %
%%%%%%%%%%%%%%%%%%%%%%%%%%%%%%%%%%%%%%%%%%%%%%
\subsubsection{Morse potential}
\label{sec:morse}

The system has finitely many discrete eigenstates
$0\le n\le n_\text{max}=[h]'$ in the specified parameter range
($[a]'$ denotes the greatest integer not exceeding and not equal to $a$):
\begin{align*}
  &U(x;h,\mu)=\mu^2e^{2x}-\mu(2h+1)e^x+h^2,\quad
  x_1=-\infty,\quad x_2=\infty,\quad h,\mu>0,\\
  &\mathcal{E}_n(h,\mu)=h^2-(h-n)^2,\quad\eta(x)=e^{-x},\\
  &\phi_n(x;h,\mu)
  =e^{hx-\mu e^x}(2\mu\eta^{-1})^{-n}L_n^{(2h-2n)}(2\mu\eta^{-1}),\quad
  h_n(h,\mu)=\frac{\Gamma(2h-n+1)}{(2\mu)^{2h}n!\,2(h-n)}.
\end{align*}
For $n>h$, the overshoot eigenfunctions provide type II seed solutions.

%%%%%%%%%%%%%%%%%%%%%%%%%%%%%%%%%%%%%%%%%%%%%%
%                                            %
% 3.2.2. Rosen-Morse potential               %
%                                            %
%%%%%%%%%%%%%%%%%%%%%%%%%%%%%%%%%%%%%%%%%%%%%%
\subsubsection{Rosen-Morse potential}
\label{sec:RM}

The system has finitely many discrete eigenstates
$0\le n\le n_\text{max}=[h-\sqrt{\mu}\,]'$ in the specified parameter range:
\begin{align*}
  &U(x;h,\mu)=-\frac{h(h+1)}{\cosh^2x}+2\mu\tanh x+h^2+\frac{\mu^2}{h^2},\quad
  x_1=-\infty,\quad x_2=\infty,\quad h>\sqrt{\mu}>0,\\
  &\mathcal{E}_n(h,\mu)=h^2-(h-n)^2+\frac{\mu^2}{h^2}-\frac{\mu^2}{(h-n)^2},
  \quad\eta(x)=\tanh x,\n
  &\phi_n(x;h,\mu)=e^{-\frac{\mu}{h-n}x}(\cosh x)^{-h+n}
  P_n\bigl(\eta(x);h,\mu\bigr),\n
  &P_n(\eta;h,\mu)=P_n^{(\alpha_n,\beta_n)}(\eta),\quad
  \alpha_n=h-n+\frac{\mu}{h-n},\ \ \beta_n=h-n-\frac{\mu}{h-n},\\
  &h_n(h,\mu)=\frac{2^{2h-2n}(h-n)
  \Gamma(h+\frac{\mu}{h-n}+1)\Gamma(h-\frac{\mu}{h-n}+1)}
  {n!\,\bigl((h-n)^2-\frac{\mu^2}{(h-n)^2}\bigr)\Gamma(2h-n+1)}.
\end{align*}
The overshoot eigenfunctions provide type II seed solutions for
$h-\sqrt{\mu}<n<h$ and type I seed solutions for $h<n<h+\sqrt{\mu}$.

%%%%%%%%%%%%%%%%%%%%%%%%%%%%%%%%%%%%%%%%%%%%%%
%                                            %
% 3.2.3. Kepler problem in hyperbolic space  %
%                                            %
%%%%%%%%%%%%%%%%%%%%%%%%%%%%%%%%%%%%%%%%%%%%%%
\subsubsection{Kepler problem in hyperbolic space}
\label{sec:Kh}

This potential is also called Eckart potential.
It has finitely many discrete eigenstates
$0\le n\le n_\text{max}=[\sqrt{\mu}-g]'$ in the specified parameter range:
\begin{align*}
  &U(x;g,\mu)=\frac{g(g-1)}{\sinh^2x}-2\mu\coth x+g^2+\frac{\mu^2}{g^2},\quad
  x_1=0,\quad x_2=\infty,\quad \sqrt{\mu}>g>\frac32,\\
  &\mathcal{E}_n(g,\mu)=g^2-(g+n)^2+\frac{\mu^2}{g^2}-\frac{\mu^2}{(g+n)^2},
  \quad\eta(x)=\coth x,\n
  &\phi_n(x;g,\mu)
  =e^{-\frac{\mu}{g+n}x}(\sinh x)^{g+n}P_n\bigl(\eta(x);g,\mu\bigr),\n
  &P_n(\eta;g,\mu)=P_n^{(\alpha_n,\beta_n)}(\eta),\quad
  \alpha_n=-g-n+\frac{\mu}{g+n},\ \ \beta_n=-g-n-\frac{\mu}{g+n},\\
  &h_n(g,\mu)=\frac{(g+n)\Gamma(1-g+\frac{\mu}{g+n})\Gamma(2g+n)}
  {2^{2g+2n}n!\,\bigl(\frac{\mu^2}{(g+n)^2}-(g+n)^2\bigr)
  \Gamma(g+\frac{\mu}{g+n})}.
\end{align*}
For $n>\sqrt{\mu}-g$, the overshoot eigenfunctions provide type I seed
solutions.

%%%%%%%%%%%%%%%%%%%%%%%%%%%%%%%%%%%%%%%%%%%%%%%%%%%%%%%%%
%                                                       %
% 3.2.4. hyperbolic Darboux-P\"{o}schl-Teller potential %
%                                                       %
%%%%%%%%%%%%%%%%%%%%%%%%%%%%%%%%%%%%%%%%%%%%%%%%%%%%%%%%%
\subsubsection{hyperbolic Darboux-P\"{o}schl-Teller potential}
\label{sec:hDPT}

This has finitely many discrete eigenstates
$0\le n\le n_\text{max}=[\frac{h-g}{2}]'$ in the specified parameter range:
\begin{align*}
  &U(x;g,h)=\frac{g(g-1)}{\sinh^2x}-\frac{h(h+1)}{\cosh^2 x}+(h-g)^2,\quad
  x_1=0,\quad x_2=\infty,\quad h>g>\frac32,\\
  &\mathcal{E}_n(g,h)=4n(h-g-n),\quad\eta(x)=\cosh 2x,\\
  &\phi_n(x;g,h)=(\sinh x)^g(\cosh x)^{-h}
  \,P_n^{(g-\frac12,-h-\frac12)}\bigl(\eta(x)\bigr),\\
  &h_n(g,h)=\frac{\Gamma(n+g+\frac12)\Gamma(h-g-n+1)}
  {2\,n!\,(h-g-2n)\Gamma(h-n+\frac12)}.
\end{align*}
The overshoot eigenfunctions provide type I seed solutions for $n>\frac{h-g}2$.

See \cite{os28} for polynomial extensions of known  solvable potentials
having finitely many discrete eigenfunctions.

%%%%%%%%%%%%%%%%%%%%%%%%%%%%%%%%%%%%%%%%%%%%%%%%%%%%%%
%                                                    %
% 3.2.5. seed solutions based on discrete symmetries %
%                                                    %
%%%%%%%%%%%%%%%%%%%%%%%%%%%%%%%%%%%%%%%%%%%%%%%%%%%%%%
\subsubsection{seed solutions based on discrete symmetries}
\label{sec:gvsds}

In \cite{os29} we have examined several well-known exactly solvable
potentials and shown that the discrete symmetries of
harmonic oscillator,
%radial oscillator, %% type I,II
%Darboux-P\"oschl-Teller potential, %% type I,II
%Coulomb potential plus the centrifugal barrier (C), %% type II
Kepler problem in spherical space,
Morse potential,
soliton potential,
Rosen-Morse potential,
hyperbolic symmetric top II,
%Kepler problem in hyperbolic space, %% type II
%hyperbolic Darboux-P\"oschl-Teller potential, %% type I,II
do not provide either type I or II virtual state wavefunctions which could be
used as seed solutions for state adding Abraham-Moses transformations.

For the hyperbolic Darboux-P\"oschl-Teller potential, Kepler problem in
hyperbolic space and Coulomb potential plus the centrifugal barrier,
the discrete symmetry produces type I or II virtual state wavefunctions.

Like the trigonometric Darboux-P\"oschl-Teller potential,
the hyperbolic Darboux-P\"oschl-Teller potential has type I and II
virtual state wavefunctions obtained by discrete symmetries
$h\leftrightarrow -(h+1)$, $g\leftrightarrow 1-g$ from the eigenfunctions
and they give seed solutions:
\begin{align*}
  \tilde{\phi}^{\I}_{\text{v}}(x;g,h)
  &=(\sinh x)^g(\cosh x)^{h+1}
  P_{\text{v}}^{(g-\frac12, h+\frac12)}\bigl(\eta(x)\bigr)
  \qquad(\text{v}\in\mathbb{Z}_{\ge 0}),\n
  \tilde{\mathcal{E}}^{\I}_{\text{v}}(g,h)
  &=-4(\text{v}+\tfrac12+g)(\text{v}+\tfrac12+h),\\
  \tilde{\phi}^{\II}_{\text{v}}(x;g,h)
  &=(\sinh x)^{1-g}(\cosh x)^{-h}
  P_{\text{v}}^{(\frac12-g,-h-\frac12)}\bigl(\eta(x)\bigr)
  \quad(\text{v}\in\mathbb{Z}_{\ge 0},\ \text{v}<\tfrac12(h+g-1)),\n
  \tilde{\mathcal{E}}^{\II}_{\text{v}}(g,h)
  &=-4(\text{v}+\tfrac12-g)(\text{v}+\tfrac12-h).
\end{align*}

The second example is Kepler problem in hyperbolic space. The virtual state
wavefunction is obtained by discrete symmetry $g\leftrightarrow 1-g$ from
the eigenfunction:
\begin{equation*}
  \tilde{\phi}_{\text{v}}(x;g,\mu)
  =e^{\frac{\mu}{g-\text{v}-1}x}(\sinh x)^{-g+\text{v}+1}
  P_{\text{v}}\bigl(\eta(x);1-g,\mu\bigr),\quad
  \tilde{\mathcal{E}}_{\text{v}}(g,\mu)=\mathcal{E}_{-\text{v}-1}(g,\mu).
\end{equation*}
For $g-1<\text{v}<g-1+\sqrt{\mu}$ ($\text{v}\in\mathbb{Z}_{\ge 0}$),
the above wavefunctions become type II seed solutions \cite{quesne5,os29}.

Coulomb potential plus the centrifugal barrier has infinitely many discrete
eigenstates in the specified parameter range:
\begin{align*}
  &U(x;g)=\frac{g(g-1)}{x^2}-\frac{2}{x}+\frac{1}{g^2},\quad
  x_1=0,\quad x_2=\infty,\quad g>\frac32,\\
  &\mathcal{E}_n(g)=\frac{1}{g^2}-\frac{1}{(g+n)^2},\quad
  \eta(x)=x^{-1},\\
  &\phi_n(x;g)
  =e^{-\frac{x}{g+n}}x^{g+n}
  \eta^nL_n^{(2g-1)}\bigl(\tfrac{2}{g+n}\eta^{-1}\bigr),\quad
  h_n(g)=\Bigl(\frac{g+n}{2}\Bigr)^{2g+2}
  \frac{4}{n!}\,\Gamma(2g+n).
\end{align*}
The discrete symmetry $g\leftrightarrow 1-g$ generates the type II seed
solutions $\text{v}>g-1$ ($\text{v}\in\mathbb{Z}_{\ge 0}$), \cite{grandati}:
\begin{equation*}
  \tilde{\phi}_{\text{v}}(x;g)
  =e^{\frac{x}{g-\text{v}-1}}x^{1-g+\text{v}}\eta^\text{v}
  L_\text{v}^{(1-2g)}\bigl(\tfrac{2}{1-g+\text{v}}\eta^{-1}\bigr),\quad
  \tilde{\mathcal{E}}_{\text{v}}(g)=\mathcal{E}_{-\text{v}-1}(g).
\end{equation*}

\bigskip

It is also possible to generalise the degree $n$ or $\text{v}$ to a real
number (or certain complex number with real energy) in the above overshoot
eigenfunctions or those wavefunctions obtained by discrete symmetry.

%%%%%%%%%%%%%%%%%%%%%%%%%%%%%%%%%%%%%%%%%%%%%%%%%%%%%%%%%%%%%%%
%                                                             %
%  4. Summary and Discussions                                 %
%                                                             %
%%%%%%%%%%%%%%%%%%%%%%%%%%%%%%%%%%%%%%%%%%%%%%%%%%%%%%%%%%%%%%%
\section{Summary and discussions}
\label{summary}

In order to carry out the program of Abraham-Moses \cite{A-M} to enlarge
the list of exactly solvable potentials through extensions by adding a
finite number of {\em eigenstates of arbitrary energies\/}, one needs
proper {\em seed solutions\/}.
Infinitely many seed solutions of different sorts are presented for some
well-known solvable potentials, {\em e.g.\/} the radial oscillator,
the Darboux-P\"oschl-Teller and the Morse potentials, etc.
They are the same {\em virtual state wavefunctions\/} which have produced
the multi-indexed Laguerre and Jacobi polynomials via multiple Darboux
transformations, and their straightforward generalisations.
There are two types of seed solutions, type I and II, corresponding to
the integral transformations starting from the lower and upper boundary
points, respectively.

The basic formulas of adding as well as deleting Abraham-Moses
transformations are recapitulated. They are presented purely algebraically
without the inverse scattering formulation. It is pointed out that the
multiple eigenstates addition transformations are a good example of
orthonormalisation procedures of non-normalisable vectors.

It would be a good challenge to formulate the difference equation analogues
of Abraham-Moses transformations. The theory of difference Schr\"odinger
equations is now well developed as `discrete quantum mechanics' \cite{os24},
and most of the orthogonal polynomials of Askey scheme \cite{askey,koeswart},
{\em e.g.\/} the Askey-Wilson and the $q$-Racah polynomials, are the
eigenfunctions of various solvable models \cite{os12,os13}. The discrete
analogues of various methods and results of quantum mechanics, including
the Heisenberg equation of motion \cite{os7}, the Darboux transformations
\cite{os15,gos}, and the multi-indexed Askey-Wilson and $q$-Racah polynomials
\cite{os27,os26} are already established.

%%%%%%%%%%%%%%%%%%%%%%%%%%%%%%%%%%%%%%%%%%%%%%%%%%%%%%%%%%%%%%%
%                                                             %
%  Acknowledgments                                            %
%                                                             %
%%%%%%%%%%%%%%%%%%%%%%%%%%%%%%%%%%%%%%%%%%%%%%%%%%%%%%%%%%%%%%%
\section*{Acknowledgements}
R.\,S. thanks Pei-Ming Ho, Jen-Chi Lee and Choon-Lin Ho for the hospitality at
National Center for Theoretical Sciences (North), National Taiwan University.
S.\,O. and R.\,S. are supported in part by Grant-in-Aid for Scientific Research
from the Ministry of Education, Culture, Sports, Science and Technology
(MEXT), No.25400395 and No.22540186, respectively.

\goodbreak
%%%%%%%%%%%%%%%%%%%%%%%%%%%%%%%%%%%%%%%%%%%%%%%%%%%%%%%%%%%%%%%
%                                                             %
%  References                                                 %
%                                                             %
%%%%%%%%%%%%%%%%%%%%%%%%%%%%%%%%%%%%%%%%%%%%%%%%%%%%%%%%%%%%%%%

%\goodbreak


\begin{thebibliography}{99}
% 
% for hyphenation : \hspace{0pt}

%%%%%%% classical integrable
\bibitem{infhul}
L.\,Infeld and T.\,E.\,Hull,
``The factorization method,''
Rev. Mod. Phys. {\bf 23} (1951) 21-68.

\bibitem{susyqm}
%See, for example, a review:
F.\,Cooper, A.\,Khare and U.\,Sukhatme,
``Supersymmetry and quantum mechanics,''
Phys. Rep. {\bf 251} (1995) 267-385.

\bibitem{nieto}
M.\,M.\,Nieto and L.\,M.\,Simmons,\,Jr.,
``Coherent States For General Potentials'',
Phys. Rev. Lett. {\bf 41} (1978) 207-210;
``Coherent States For General Potentials". 1. Formalism,
Phys. Rev. D {\bf 20} (1979) 1321-1331;
2. Confining One-Dimensional Examples,
Phys. Rev. D {\bf 20} (1979) 1332-1341;
3. Nonconfining One-Dimensional Examples,
Phys. Rev. D {\bf 20} (1979) 1342-1350.

%%%%%%% Darboux-Crum
\bibitem{darb}
G.\,Darboux,
{\it Th\'eorie g\'en\'erale des surfaces}
vol 2 (1888) Gauthier-Villars, Paris.

\bibitem{crum}
M.\,M.\,Crum,
``Associated Sturm-Liouville systems,"
Quart. J. Math. Oxford Ser. (2) {\bf 6} (1955) 121-127,
{\tt arXiv:physics/9908019}.

%%%%%%% A-M
\bibitem{A-M}
P.\,B.\,Abraham and H.\,E.\,Moses,
``Changes in potentials due to changes in the point spectrum:
Anharmonic oscillators with exact solutions,"
Phys. Rev. {\bf A22} (1980) 1333-1340.

%%%%%%% multi-index
\bibitem{os25}
S.\,Odake and R.\,Sasaki,
``Exactly Solvable Quantum Mechanics and Infinite Families of
Multi-indexed Orthogonal Polynomials,"
Phys. Lett. {\bf B702} (2011) 164-170,
{\tt arXiv:\hspace{0pt}1105.0508[math-ph]}.

\bibitem{gomez3}
D.\,G\'{o}mez-Ullate, N.\,Kamran and R.\,Milson,
``Two-step Darboux transformations and exceptional Laguerre polynomials,"
J. Math. Anal. Appl. {\bf 387} (2012) 410-418,
{\tt arXiv:\hspace{0pt}1103.5724[math-ph]}.

%%%%%%% exceptionals
\bibitem{os16}
S.\,Odake and R.\,Sasaki,
``Infinitely many shape invariant potentials and new orthogonal polynomials,''
Phys. Lett. {\bf B679} (2009) 414-417,
{\tt arXiv:0906.0142[math-ph]}.

\bibitem{os19}
S.\,Odake and R.\,Sasaki,
``Another set of infinitely many exceptional ($X_{\ell}$) Laguerre
polynomials,''
Phys. Lett. {\bf B684} (2010) 173-176,
{\tt arXiv:0911.3442[math-ph]}.

\bibitem{gomez}
D.\,G\'{o}mez-Ullate, N.\,Kamran and R.\,Milson,
``An extension of Bochner's problem: exceptional invariant subspaces,''
J. Approx Theory {\bf 162} (2010) 987-1006,
{\tt arXiv:0805.\hspace{0pt}3376[math-ph]};
``An extended class of orthogonal polynomials defined by a
Sturm-Liouville problem,''
J. Math. Anal. Appl. {\bf 359} (2009) 352-367,
{\tt arXiv:0807.3939[math-\hspace{0pt}ph]}.

\bibitem{quesne}
C.\,Quesne,
``Exceptional orthogonal polynomials, exactly solvable potentials and
supersymmetry,''
J. Phys. {\bf A41} (2008) 392001 (6pp),
{\tt arXiv:0807.4087[quant-ph]}.

\bibitem{bqr}
B.\,Bagchi, C.\,Quesne and R.\,Roychoudhury,
``Isospectrality of conventional and new extended potentials,
second-order supersymmetry and role of PT symmetry,"
Pramana J. Phys. {\bf 73} (2009) 337-347,
{\tt arXiv:0812.1488[quant-ph]}.

\bibitem{quesne2}
C.\,Quesne,
``Solvable rational potentials and exceptional orthogonal polynomials
in supersymmetric quantum mechanics,"
SIGMA {\bf 5} (2009) 084 (24pp),
{\tt arXiv:0906.2331\hspace{0pt}[math-ph]}.

\bibitem{hos}
C.-L.\,Ho, S.\,Odake and R.\,Sasaki,
``Properties of the exceptional ($X_{\ell}$) Laguerre and Jacobi polynomials,''
SIGMA {\bf 7} (2011) 107 (24pp),
{\tt arXiv:0912.5447[math-ph]}.

%%%%%%% A-M related
\bibitem{AM-rel}
M.\,M.\,Nieto,
``Relationship between supersymmetry and the inverse method in quantum
mechanics,"
Phys. Lett. {\bf 145B} (1984) 208-210;
B.\,Mielnik,
``Factorization method and new potentials with the oscillator spectrum,"
J. Math. Phys. {\bf 25} (1984) 3387-3389;
C.\,V.\,Sukumar,
``Supersymmetric quantum mechanics of one-dimensional systems,"
J. Phys. {\bf A18} (1985) 2917-2936.

\bibitem{pursey}
M.\,Luban and D.\,L.\,Pursey,
``New Schrodinger equations for old: Inequivalence of the Darboux and
Abraham-Moses constructions,"
Phys. Rev. {\bf D33} (1986) 431-436;
D.\,L.\,Pursey,
``New families of isospectral Hamiltonians,"
Phys. Rev. {\bf D33} (1986) 1048-1055.

\bibitem{trl}
L.\,Trlifaj,
``The Darboux and Abraham-Moses transformations of the one-dimensional
periodic Schr\"odinger equation and inverse problems,"
Inverse Problems {\bf 5} (1989) 1145-1155.

\bibitem{schn-leeb}
W.\,A.\,Schnitzer and H.\,Leeb, 
``Generalized Darboux transformations: classification of inverse scattering
methods for the radial Schr\"odinger equation,"
J. Phys. {\bf A27} (1994) 2605-2614.

\bibitem{sam}
B.\,F.\,Samsonov,
``On the equivalence of the integral and the differential exact solution
generation methods for the one-dimensional Schr\"odinger equation,"
J. Phys. {\bf A28} (1995) 6989-6998.

%%%%%%% inverse
\bibitem{inv}
I.\,M.\,Gel'fand and B.\,M.\,Levitan,
``On the determination of a differential equation from its spectral function,"
(Russian)
Izvestiya Akad. Nauk SSSR. Ser. Mat. {\bf 15} (1951) 309-360
(Amer. Math. Soc. Transl. Ser.2 {\bf 1} (1955) 253-304);
K.\,Chadan and P.\,C.\,Sabatier,
``Inverse problems in quantum scattering theory,"
Springer Verlag, New York (1977).

\bibitem{adler}
M.\,G.\,Krein,
``On continuous analogue of a formula of Christoffel from the theory
of orthogonal polynomials," (Russian)
Doklady Acad. Nauk. CCCP {\bf 113} (1957) 970-973;
V.\,\'E.\,Adler,
``A modification of Crum's method,''
Theor. Math. Phys. {\bf 101} (1994) 1381-1386.

\bibitem{os28}
S.\,Odake and R.\,Sasaki,
``Extensions of solvable potentials with finitely many discrete eigenstates,"
J. Phys. {\bf A46} (2013) 235205 (15pp),
{\tt arXiv:1301.3980[math-ph]}.

\bibitem{os29}
S.\,Odake and R.\,Sasaki,
``Krein-Adler transformations for shape-invariant potentials and pseudo
virtual states,"
J. Phys. {\bf A46} (2013) 245201 (24pp),
{\tt arXiv:1212.6595[math-\hspace{0pt}ph]}.

\bibitem{os21}
S.\,Odake and R.\,Sasaki,
``A new family of shape invariantly deformed Darboux-P\"oschl-Teller
potentials with continuous $\ell$,"
J. Phys. {\bf A44} (2011) 195203 (14pp),
{\tt arXiv:1007.\hspace{0pt}3800[math-ph]}.

\bibitem{junkroy}
G.\,Junker and P.\,Roy,
``Conditionally exactly solvable problems and non-linear algebras,"
Phys. Lett. {\bf A 232} (1997) 155-161;
G.\,Junker and P.\,Roy,
``Conditionally Exactly Solvable Potentials: A Supersymmetric Construction
Method,"
Ann. Phys. {\bf 270} (1998) 155-177.

%%%%%%% sinusoidal
\bibitem{os7}
S.\,Odake and R.\,Sasaki,
``Unified theory of annihilation-creation operators for solvable
(`discrete') quantum mechanics,''
J. Math. Phys. {\bf 47} (2006) 102102 (33pp),
{\tt arXiv:\hspace{0pt}quant-ph/0605215};
``Exact solution in the Heisenberg picture and annihilation-creation
operators,"
Phys. Lett. {\bf B641} (2006) 112-117,
{\tt arXiv:quant-ph/0605221}.

%%%%%%% Morse etc
\bibitem{quesne4}
C.\,Quesne,
``Revisiting (quasi-)exactly solvable rational extensions of the Morse
potential,"
Int. J. Mod. Phys. {\bf A 27} (2012) 1250073 (18pp),
{\tt arXiv:1203.1812[math-ph]}.

\bibitem{quesne5}
C.\,Quesne,
``Novel Enlarged Shape Invariance Property and Exactly Solvable Rational
Extensions of the Rosen-Morse II and Eckart Potentials,"
SIGMA {\bf 8} (2012) 080 (19pp),
{\tt arXiv:1208.6165[math-ph]}.

\bibitem{grandati2}
Y.\,Grandati, 
``New rational extensions of solvable potentials with finite bound state
spectrum,''
Phys. Lett. {\bf A376} (2012) 2866-2872,
{\tt arXiv:1203.4149[math-ph]}.

\bibitem{grandati}
Y.\,Grandati,
``Solvable rational extensions of the isotonic oscillator,''
{\tt arXiv:1101.0055\hspace{0pt}[math-ph]};
``Solvable rational extensions of the Morse and Kepler-Coulomb potentials,"
J. Math. Phys. {\bf 52} (2011) 103505 (12pp),
{\tt arXiv:1103.5023[math-ph]}.

\bibitem{os24}
S.\,Odake and R.\,Sasaki,
``Discrete quantum mechanics," (Topical Review)
J. Phys. {\bf A44} (2011) 353001 (47pp),
{\tt arXiv:1104.0473[math-ph]}.

\bibitem{askey}
G.\,E.\,Andrews, R.\,Askey and R.\,Roy,
{\it Special Functions\/},
vol. 71 of Encyclopedia of mathematics and its applications,
Cambridge Univ. Press, Cambridge, (1999).

\bibitem{koeswart}
R.\,Koekoek and R.\,F.\,Swarttouw,
``The Askey-scheme of hypergeometric orthogonal polynomials and
its $q$-analogue,''
{\tt arXiv:math.CA/9602214}.

\bibitem{os12}
S.\,Odake and R.\,Sasaki,
``Orthogonal Polynomials from Hermitian Matrices,"
J. Math. Phys. {\bf 49} (2008) 053503 (43pp),
{\tt arXiv:0712.4106[math.CA]}.

\bibitem{os13}
S.\,Odake and R.\,Sasaki,
``Exactly solvable `discrete' quantum mechanics; shape invariance,
Heisenberg solutions, annihilation-creation operators and coherent states,"
Prog. Theor. Phys. {\bf 119} (2008) 663-700,
{\tt arXiv:0802.1075[quant-ph]}.

\bibitem{os15}
S.\,Odake and R.\,Sasaki,
``Crum's Theorem for `Discrete' Quantum Mechanics,"
Prog. Theor. Phys. {\bf 122} (2009) 1067-1079,
{\tt arXiv:0902.2593[math-ph]}.

\bibitem{gos}
Leonor Garc\'ia-Guti\'errez, S.\,Odake and R.\,Sasaki,
``Modification of Crum's Theorem for `Discrete' Quantum Mechanics,"
Prog. Theor. Phys. {\bf 124} (2010) 1-24,
{\tt arXiv:1004.0289\hspace{0pt}[math-ph]}.

%%%%%%% discrete
\bibitem{os27}
S.\,Odake and R.\,Sasaki,
``Multi-indexed Wilson and Askey-Wilson polynomials,"
J. Phys. {\bf A46} (2013) 045204 (22pp),
{\tt arXiv:1207.5584[math-ph]}.

\bibitem{os26}
S.\,Odake and R.\,Sasaki, 
``Multi-indexed ($q$-)Racah polynomials,"
J. Phys. {\bf A45} (2012) 385201 (21pp),
{\tt arXiv:1203.5868[math-ph]}.

\end{thebibliography}
\end{document}